%% file: nnc-conv.tex
\documentclass{llncs}

\usepackage{amsmath}
\usepackage{amssymb}
\usepackage{booktabs}

\makeatletter
\newenvironment{delayedproof}[1][\proofname]
  {\par\normalfont \topsep6\p@ \@plus6\p@\relax
  \trivlist
  \item[\hskip\labelsep\textbf{Proof of #1\@addpunct{.}}]\ignorespaces}
  {\endtrivlist\@endpefalse}
\makeatother

\usepackage{pdftricks}
\begin{psinputs}
  \usepackage{pstricks}
\end{psinputs}
\usepackage[pdf]{pstricks}

\usepackage{float}
\usepackage[hypertexnames=false]{hyperref}
\usepackage{algorithm}
\usepackage[noend]{algpseudocode}
\floatname{algorithm}{Pseudocode}

\usepackage{cleveref}

\usepackage{tikz}
\usetikzlibrary{matrix,shapes,arrows,positioning,chains}

\usepackage{hhline}
\usepackage{multirow}

\include{macros}

\begin{document}

\input{frontmatter}

\section{Introduction}
\label{sec:intro}
\input{intro}

\section{Preliminaries}
\label{sec:prelims}
\input{prelims}

\section{NNC Polyhedra as Closed Polyhedra}
\label{sec:old-repr}
\input{old-repr}

\section{Direct Representations for NNC Polyhedra}
\label{sec:new-repr}
\input{new-rep}

\section{The New Conversion Algorithm}
\label{sec:conv}
\input{conv}

\section{Duality}
\label{sec:duality}
\input{duality}

\section{Experimental Evaluation}
\label{sec:exp-eval}
\input{exp-eval}

\section{Conclusion}
\label{sec:concl}
\input{concl}


\newcommand{\noopsort}[1]{}\hyphenation{ Ba-gna-ra Bie-li-ko-va Bruy-noo-ghe
  Common-Loops DeMich-iel Dober-kat Di-par-ti-men-to Er-vier Fa-la-schi
  Fell-eisen Gam-ma Gem-Stone Glan-ville Gold-in Goos-sens Graph-Trace
  Grim-shaw Her-men-e-gil-do Hoeks-ma Hor-o-witz Kam-i-ko Kenn-e-dy Kess-ler
  Lisp-edit Lu-ba-chev-sky Ma-te-ma-ti-ca Nich-o-las Obern-dorf Ohsen-doth
  Par-log Para-sight Pega-Sys Pren-tice Pu-ru-sho-tha-man Ra-guid-eau Rich-ard
  Roe-ver Ros-en-krantz Ru-dolph SIG-OA SIG-PLAN SIG-SOFT SMALL-TALK Schee-vel
  Schlotz-hauer Schwartz-bach Sieg-fried Small-talk Spring-er Stroh-meier
  Thing-Lab Zhong-xiu Zac-ca-gni-ni Zaf-fa-nel-la Zo-lo }

\appendix
\section{Appendix}
\label{sec:appendix}
\input{appendix}

\end{document}

%% file: macros.tex
\newcommand*{\cC}{\ensuremath{\mathcal{C}}}

\newcommand*{\cF}{\ensuremath{\mathcal{F}}}
\newcommand*{\cG}{\ensuremath{\mathcal{G}}}

\newcommand*{\cP}{\ensuremath{\mathcal{P}}}
\newcommand*{\cQ}{\ensuremath{\mathcal{Q}}}
\newcommand*{\cR}{\ensuremath{\mathcal{R}}}

\newcommand*{\fund}[3]{\mathord{#1}\colon#2\rightarrow#3}

\newcommand*{\comp}{\circ}

\newcommand{\defrel}[1]{\mathrel{\buildrel \mathrm{def} \over {#1}}}
\newcommand{\defeq}{\defrel{=}}

\providecommand*{\Rset}{\mathbb{R}}            
\providecommand*{\nonnegRset}{\mathbb{R}_{\scriptscriptstyle{+}}}
\providecommand*{\CPset}{\mathbb{CP}}          
\providecommand*{\Pset}{\mathbb{P}}            
\providecommand*{\polyhull}{\uplus}

\newcommand*{\vect}[1]{\mathchoice{\mbox{\boldmath$\displaystyle#1$}}
{\mbox{\boldmath$\textstyle#1$}}
{\mbox{\boldmath$\scriptstyle#1$}}
{\mbox{\boldmath$\scriptscriptstyle#1$}}} 
\newcommand*{\transpose}{{\scriptscriptstyle\mathrm{T}}}

\newcommand*{\sseq}{\subseteq}

\newcommand*{\Sseq}{\supseteq}

\newcommand*{\sqsseq}{\sqsubseteq}
\newcommand*{\sqsslt}{\sqsubset}

\newcommand*{\wpup}{\mathop{\wp_{\uparrow}}\nolimits}

\newcommand*{\union}{\mathrel{\cup}}
\newcommand*{\bigunion}{\bigcup}

\newcommand*{\inters}{\mathrel{\cap}}
\newcommand*{\setdiff}{\mathrel{\setminus}}

\newcommand{\sset}[2]{{\renewcommand{\arraystretch}{1.2}
                      \left\{\,#1 \,\left|\,
                               \begin{array}{@{}l@{}}#2\end{array}
                      \right.   \,\right\}}}

\newcommand*{\cone}[1]{\mathbb{#1}}
\newcommand*{\conesem}[1]{\left[\![{#1}\right]\!]_{\xi}}
\newcommand*{\epssem}[1]{\left[\![{#1}\right]\!]_{\epsilon}}

\newcommand{\st}{\mathrel{.}}
\newcommand{\itc}{\mathrel{:}}

\newcommand*{\cl}{\mathop{\mathrm{cl}}\nolimits}
\newcommand*{\con}{\mathop{\mathrm{con}}\nolimits}
\newcommand*{\gen}{\mathop{\mathrm{gen}}\nolimits}

\newcommand*{\fullgen}{\mathop{\mathrm{full.gen}}\nolimits}

\newcommand*{\relint}{\mathop{\mathrm{relint}}\nolimits}

\newcommand*{\satinter}{\mathop{\mathrm{sat.inter}}\nolimits}

\newcommand*{\proj}{\mathop{\mathrm{proj}}\nolimits}

\newcommand*{\suppcl}{\mathop{\mathrm{supp.cl}}\nolimits}

\newcommand*{\skel}{\mathop{\mathrm{skel}}\nolimits}

\newcommand*{\NSset}{\mathbb{NS}}

\newcommand*{\ns}{\mathit{ns}}

\newcommand*{\SK}{\ensuremath{\mathcal{SK}}}
\newcommand*{\NS}{\ensuremath{\mathit{NS}}}
\newcommand*{\SP}{\mathit{SP}}

\newcommand*{\SC}{\mathit{SC}}

\newcommand*{\NNCFaces}{\mathit{nncFaces}}
\newcommand*{\CFaces}{\mathit{cFaces}}
\newcommand*{\GFaces}{\mathit{gFaces}}

\newcommand*{\upcl}{\mathop{\uparrow}\nolimits}

\newcommand{\relop}{\mathrel{\bowtie}}

\newcommand*{\convstep}[1]{\xrightarrow{#1}}
\newcommand*{\combine}[1]{\mathop{\mathrm{combine}_{#1}}\nolimits}
\newcommand*{\combineadj}[1]{\mathop{\mathrm{combine\_adj}_{#1}}\nolimits}
\newcommand*{\adj}{\mathop{\mathrm{adjacent}}\nolimits}

\newcommand*{\pbecomecp}
            {\mathop{\mathrm{points\_become\_closure\_points}}\nolimits}

\newcommand*{\mytablesize}{\footnotesize}

\newcommand*{\Cplusplus}{{C\nolinebreak[4]\hspace{-.05em}\raisebox{.4ex}{\tiny\bf ++}}}

%% file: frontmatter.tex
\title{A Conversion Procedure for NNC Polyhedra}

\author{Anna Becchi, Enea Zaffanella}
\authorrunning{A. Becchi, E. Zaffanella}

\institute{Department of Mathematical, Physical and Computer Sciences\\
  University of Parma, Italy\\
  \email{anna.becchi@studenti.unipr.it\\
    enea.zaffanella@unipr.it}
}

\maketitle

\begin{abstract}
  We present an alternative Double Description representation for the domain
  of NNC (not necessarily topologically closed) polyhedra,
  together with the corresponding Chernikova-like conversion procedure.
  The representation differs from the ones adopted in the currently
  available implementations of the Double Description method in that
  it uses no slack variable at all: this new approach provides a solution
  to a few technical issues caused by the encoding of an NNC polyhedron
  as a closed polyhedron in a higher dimension space.
  A preliminary experimental evaluation shows that the new conversion
  algorithm is able to achieve significant efficiency improvements
  with respect to state-of-the-art implementations.
\end{abstract}

%% file: intro.tex
The Double Description (DD) method~\cite{MotzkinRTT53}
allows for the representation and manipulation of convex polyhedra
by using two different geometric representations:
one based on a finite collection of \emph{constraints},
the other based on a finite collection of \emph{generators}.
Starting from any one of these representations,
the other can be derived by application of a conversion
procedure~\cite{Chernikova64,Chernikova65,Chernikova68},
thereby obtaining a DD pair;
the procedure allows for the identification and removal
of redundant elements from both representations,
yielding a DD pair in minimal form;
moreover, it is incremental, allowing for capitalizing
on the work already done when new constraints and/or generators
need to be added to an input DD pair.

The DD method lies at the foundation of several software libraries and tools.
The following is an incomplete list of available implementations:
\begin{itemize}
\item
cdd
(\url{www.inf.ethz.ch/personal/fukudak/cdd\_home});
\item
PolyLib (\url{http://icps.u-strasbg.fr/PolyLib});
\item
NewPolka, part of Apron
(\url{http://apron.cri.ensmp.fr/library});
\item
Parma Polyhedra Library
(\url{http://bugseng.com/products/ppl});
\item
4ti2 (\url{www.4ti2.de});
\item
Skeleton (\url{www.uic.unn.ru/~zny/skeleton});
\item
Addibit
(\url{www.informatik.uni-bremen.de/agbs/bgenov/addibit}).
\end{itemize}

Despite the intrinsic exponential complexity of the conversion procedure,
these implementations turn out to be surprisingly effective in many contexts.
As a consequence, the range of applicability of the DD method keeps widening,
also due to several incremental improvements in the efficiency
of the most critical processing phases
\cite{FukudaP96,Genov14PhD,LeVerge92,Zolotykh12}.
Quoting from~\cite{FukudaP96}:
\begin{quote}
  The double description method is a simple and useful algorithm
  [\dots]
  despite the fact that we can hardly state any interesting theorems
  on its time and space complexities.
\end{quote}
Implementations of the DD method are actively used,
either directly or indirectly, in several research fields,
with applications as diverse as
bioinformatics~\cite{TerzerS08,TerzerS09},
computational geometry~\cite{4ti2,AssarfGHJLPR17},
analysis of analog and hybrid systems
\cite{BenerecettiFM13,Frehse08,HalbwachsPR94,HalbwachsPR97},
automatic parallelization~\cite{Bastoul04,Pop06},
scheduling~\cite{DooseM05},
static analysis of software
\cite{BagnaraHZ09TCS,ColonS01,CousotH78,Ellenbogen04th,Gopan07th,HenryMM12}.

In the classical setting, the DD method is meant to compute
geometric representations for \emph{topologically closed} polyhedra
in an $n$-dimensional vector space.
However, there are applications requiring the ability to also deal
with linear \emph{strict} inequality constraints, leading to the
definition of \emph{not necessarily closed} (NNC) polyhedra.
For example, this is the case for some of the analysis tools
developed for the verification of
hybrid systems~\cite{BenerecettiFM13,Frehse08,HalbwachsPR94,HalbwachsPR97};
other examples of the use of NNC polyhedra include static analysis
tools such as Pagai~\cite{HenryMM12}, where strict inequality
constraints are used to model the semantics of conditional tests
acting on program variables of floating point type,
as well as the automatic discovery of ranking functions~\cite{ColonS01}
for proving the termination of program fragments.

The few DD method implementations providing support for
NNC polyhedra are all based on an \emph{indirect} representation of the
strict inequalities, which are encoded by adding an additional
space dimension playing the role of a slack variable.
The main advantage of this approach is the possibility of reusing,
almost unchanged, all of the well-studied algorithms and optimizations
that have been developed for the classical case of closed polyhedra
\cite{BagnaraHZ05FAC,BagnaraRZH02,HalbwachsPR94,HalbwachsPR97}.
However, the addition of a slack variable carries with itself
an obvious overhead, as well as a few technical issues.

In this paper, we pursue a different approach for the handling
of NNC polyhedra in the DD method.
Namely, we specify a \emph{direct} representation,
dispensing with the need of the slack variable.
The main insight of this new approach is the separation
of the (constraints or generators) geometric representation
into two components, the skeleton and the non-skeleton
of the representation, playing quite different roles:
while keeping a geometric encoding for the skeleton component,
we will adopt a combinatorial encoding for the non-skeleton one.
For this new representation, we propose the corresponding variant
of the Chernikova's conversion procedure,
where both components are handled by respective
processing phases, so as to take advantage of their peculiarities.
In particular, we develop \emph{ad hoc} functions and
procedures for the combinatorial non-skeleton part.

The new representation and conversion procedure, in principle,
can be integrated into any of the available implementations of
the DD method. Our implementation and experimental evaluation,
conducted in the context of the Parma Polyhedra Library,
show that the new algorithm, while computing the correct results for
all of the considered tests, achieves impressive efficiency improvements
with respect to the implementation based on the slack variable.

The paper is structured as follows.
Section~\ref{sec:prelims}, after introducing the required
notation and terminology, briefly describes the Double Description
method for the representation of closed polyhedra,
also sketching the Chernikova's conversion algorithm.
Section~\ref{sec:old-repr} summarizes the encoding of NNC polyhedra
into closed polyhedra based on the addition of a slack variable,
highlighting a few technical issues.
Section~\ref{sec:new-repr}
proposes the new representation for NNC polyhedra,
which uses no slack variable and distinguishes between
a geometric and a combinatorial component.
Section~\ref{sec:conv} is devoted to the extension of
the Chernikova's conversion algorithm to the case of NNC polyhedra
adopting this new representation.
Section~\ref{sec:duality} shows how, by applying duality arguments,
all the concepts and results presented in Sections~\ref{sec:new-repr}
and~\ref{sec:conv} for the case of generators can be generalized
to also deal with the case of constraints.
Section~\ref{sec:exp-eval} reports the results obtained
by the experimental evaluation of the new algorithm.
We conclude in Section~\ref{sec:concl}.
Proof sketches for the stated results can be found
in Appendix~\ref{sec:appendix}.

This paper is a revision and extension of~\cite{Becchi17th},
where the new representation was introduced and
the conversion procedure from constraints to generators
was initially proposed and experimentally evaluated.

%% file: prelims.tex
We assume some familiarity with the basic notions of
lattice theory~\cite{Birkhoff67}.

For a lattice $\langle L, \sqsseq, \bot, \top, \sqcap, \sqcup \rangle$,
an element $a \in L$ is an \emph{atom} if $\bot \sqsslt a$
and there exists no element $b \in L$ such that $\bot \sqsslt b \sqsslt a$.
The lattice $L$ is said to be \emph{atomistic} if every element of $L$
can be obtained as the join of a set of atoms.
For $S \sseq L$, the \emph{upward closure} of $S$ is defined as
\(
  \upcl S \defeq \{\, x \in L \mid \exists s \in S \st s \sqsseq x \,\}
\).
The set $S \sseq L$ is \emph{upward closed} if $S = \upcl S$;
we denote by $\wpup(L)$ the set of all
the upward closed subsets of $L$.
For $x \in L$, $\upcl x$ is a shorthand for $\upcl \{ x \}$.
The notation for \emph{downward closure} is similar.

Given two posets
$\langle L, \sqsseq \rangle$ and $\langle L^\sharp, \sqsseq^\sharp \rangle$
and two monotonic functions
$\fund{\alpha}{L}{L^\sharp}$ and $\fund{\gamma}{L^\sharp}{L}$,
the pair $(\alpha, \gamma)$ is a \emph{Galois connection}~\cite{CousotC79}
(between $L$ and $L^\sharp$) if
\[
  \forall x \in L, x^\sharp \in L^\sharp
    \itc
      \alpha(x) \sqsseq^\sharp x^\sharp
        \Leftrightarrow
          x \sqsseq \gamma(x^\sharp).
\]

We write $\Rset^n$ to denote the Euclidean topological space
of dimension $n > 0$ and $\nonnegRset$ for the set of non-negative reals;
for $S \sseq \Rset^n$, $\cl(S)$ and $\relint(S)$
denote the topological closure and the relative interior of $S$,
respectively.
The scalar product of two vectors $\vect{a}_1, \vect{a}_2 \in \Rset^n$
is denoted by $\vect{a}_1^\transpose \vect{a}_2$.
For each vector $\vect{a} \in \Rset^n$, where $\vect{a} \neq \vect{0}$,
and scalar $b \in \Rset$,
the linear non-strict inequality constraint
$\beta = ( \vect{a}^\transpose \vect{x} \geq b )$
defines a closed affine half-space of $\Rset^n$;
similarly, the linear equality constraint
$\beta = ( \vect{a}^\transpose \vect{x} = b )$
defines an affine hyperplane of $\Rset^n$.

A topologically closed convex polyhedron (for short, closed polyhedron)
is defined as the set of solutions of a finite system $\cC$
of linear non-strict inequality and linear equality constraints;
namely, $\cP = \con(\cC)$ where
\[
  \con(\cC)
    \defeq
      \bigl\{\,
        \vect{p} \in \Rset^n
      \bigm|
        \forall \beta = ( \vect{a}^\transpose \vect{x} \relop b) \in \cC,
        \mathord{\relop} \in \{ \mathord{\geq}, \mathord{=} \}
          \st
            \vect{a}^\transpose \vect{p} \relop b
      \,\bigr\}.
\]

A vector $\vect{r} \in \Rset^n$ such that $\vect{r} \neq \vect{0}$
is a \emph{ray} of a non-empty polyhedron
$\cP \sseq \Rset^n$ if, for every point $\vect{p} \in \cP$
and every non-negative scalar $\rho \in \nonnegRset$,
it holds $\vect{p} + \rho \vect{r} \in \cP$.
The empty polyhedron has no rays.
If both $\vect{r}$ and $-\vect{r}$ are rays of $\cP$, then
we say that $\vect{r}$ is a \emph{line} of $\cP$.
By Minkowski and Weyl theorems~\cite{StoerW70},
the set $\cP \sseq \Rset^n$ is a closed polyhedron if and only if
there exist finite sets $L, R, P \sseq \Rset^n$
of cardinality $\ell$, $r$ and $p$, respectively,
such that $\vect{0} \notin (L \union R)$ and
$\cP = \gen\bigl(\langle L, R, P \rangle\bigr)$,
where
\[
  \gen\bigl( \langle L, R, P \rangle \bigr)
      \defeq
        \bigl\{\,
          L \vect{\lambda} + R \vect{\rho} + P \vect{\pi} \in \Rset^n
        \bigm|
          \vect{\lambda} \in \Rset^{\ell},
          \vect{\rho} \in \nonnegRset^r,
          \vect{\pi} \in \nonnegRset^p,
          \textstyle{\sum_{i=1}^p \pi_i = 1}
        \,\bigr\}.
\]
When $\cP \neq \emptyset$, we say that $\cP$ is described by
the \emph{generator system} $\cG = \langle L, R, P \rangle$.
In the following, we will abuse notation
by adopting the usual set operator and relation symbols
to denote the corresponding component-wise extensions
on generator systems.
For instance,
for $\cG = \langle L, R, P \rangle$
and $\cG' = \langle L', R', P' \rangle$,
we will write $\cG \sseq \cG'$ to mean
$L \sseq L'$, $R \sseq R'$ and $P \sseq P'$;
similarly, we may write $\wp(\cG)$ to denote the set of
all generator systems $\cG'$ such that $\cG' \sseq \cG$.

The Double Description method due to Motzkin et al.~\cite{MotzkinRTT53},
by exploiting the duality principle, allows to combine
the constraints and the generators of a polyhedron $\cP$
into a DD pair $(\cC, \cG)$: a \emph{conversion} procedure is used
to obtain each description starting from the other one,
also removing the redundant elements.
For presentation purposes, we focus on the conversion from constraints
to generators; the conversion from generators to constraints works
in the same way, using duality to switch the roles of constraints
and generators.

The conversion procedure starts from a DD pair
$(\cC_0, \cG_0)$ representing the whole vector space
and adds, one at a time,
the elements of the input constraint system
$\cC = \{ \beta_0, \dots, \beta_m \}$,
producing a sequence of DD pairs
$\bigl\{(\cC_k, \cG_k)\bigr\}_{0 \leq k \leq m+1}$
representing the polyhedra
\[
  \Rset^n = \cP_0
    \convstep{\beta_0} \dots
    \convstep{\beta_{k-1}} \cP_k
    \convstep{\beta_k} \cP_{k+1}
    \convstep{\beta_{k+1}} \dots
    \convstep{\beta_m} \cP_{m+1} = \cP.
\]
At each iteration,
when adding the constraint $\beta_k$ to polyhedron $\cP_k = \gen(\cG_k)$,
the generator system $\cG_k$ is partitioned into
the three components $\cG^{+}_k$, $\cG^{0}_k$, $\cG^{-}_k$,
according to the sign of the scalar products of the generators with $\beta_k$
(those in $\cG^{0}_k$ are the \emph{saturators} of $\beta_k$);
the new generator system for polyhedron $\cP_{k+1}$
is computed as
$\cG_{k+1} \defeq \cG^{+}_k \union \cG^{0}_k \union \cG^{\star}_k$,
where
\begin{align*}
  \cG^{\star}_k
    &= \combineadj{\beta_k}(\cG^{+}_k, \cG^{-}_k) \\
    &\defeq
      \bigl\{\,
        \combine{\beta_k}(g^{+}, g^{-})
      \bigm|
        g^{+} \in \cG^{+}_k, g^{-} \in \cG^{-}_k,
        \adj_{\cP_k}(g^{+}, g^{-})
      \,\bigr\}.
\end{align*}
Function `$\mathord{\combine{\beta_k}}$' computes a linear combination
of its arguments, yielding a generator that saturates the constraint $\beta_k$;
predicate `$\adj_{\cP_k}$' is used to discard those pairs
of generators that are not \emph{adjacent} in $\cP_k$
(since these would only produce redundant generators).

The conversion procedure is usually followed by a \emph{simplification}
step, where the DD pair is modified, without affecting the represented
polyhedron, so as to achieve some form of minimality.
For instance, the implicit linear equality constraints
(encoded by non-strict inequalities) are detected and represented
explicitly; similarly, rays are combined to produce lines.
We will not provide a formalization of these details,
assuming anyway that these simplifications are implicitly taken
into proper account when needed.

Similarly, it is worth noting that the one sketched above
is a high level description of the conversion procedure;
at the implementation level, each closed polyhedron
$\cP \sseq \Rset^n$ is mapped, by \emph{homogenization},
into a (topologically closed) convex polyhedral cone
$\cone{C} \sseq \Rset^{n+1}$.
This process associates a new space dimension, usually denoted as $\xi$,
to the inhomogeneous term of constraints; the new space dimension
is constrained to only assume non-negative values,
i.e., the \emph{positivity constraint} $\xi \geq 0$
is added to the constraint representation of the polyhedral cone.
When reinterpreted in the $n$ dimensional vector space,
this constraint can be read as the tautology $1 \geq 0$.
The inverse map from a convex polyhedral cone $\cone{C}$
to the represented convex polyhedron $\cP$
is obtained by only considering the points of the cone
having a strictly positive coordinate for the $\xi$ dimension:
\[
\cP = \conesem{\cone{C}}
  \defeq
    \bigl\{\,
      \vect{x} \in \Rset^n
    \bigm|
      (\vect{x}^\transpose, \xi)^\transpose \in \cone{C}, \xi > 0
    \,\bigr\}.
\]
By homogenization, all of the vertices of the convex polyhedron
are mapped into rays of the convex polyhedral cone:
this also allows for a more uniform handling of the rays and vertices,
a property which is suitably exploited in most implementations.
The rays of the convex polyhedral cone can be easily reinterpreted:
those having a zero (resp., positive) coordinate for the $\xi$
space dimension are the rays (resp., points)
of the represented polyhedron.

The set $\CPset_n$ of all closed polyhedra on the vector space $\Rset^n$,
partially ordered by set inclusion, is a lattice
\(
  \langle\,
    \CPset_n, \sseq, \emptyset, \Rset^n, \mathord{\inters}, \mathord{\polyhull}
  \,\rangle
\),
where the emptyset and $\Rset^n$ are the bottom and top elements,
the binary meet operator is set intersection and the binary join
operator `$\mathord{\polyhull}$' is the convex polyhedral hull.

A linear \emph{strict} inequality constraint
$\beta = (\vect{a}^\transpose \vect{x} > 0)$
defines an open affine half-space of $\Rset^n$.
When the constraint system $\cC$ is extended to also allow
for strict inequalities, the convex polyhedron
$\cP = \con(\cC)$ is not necessarily (topologically) closed.
The set $\Pset_n$ of all NNC polyhedra on the vector space $\Rset^n$
is a lattice
\(
  \langle\,
    \Pset_n, \sseq, \emptyset, \Rset^n, \mathord{\inters}, \mathord{\polyhull}
  \,\rangle
\)
and $\CPset_n$ is a sublattice of $\Pset_n$.

As shown in~\cite{BagnaraHZ05FAC,BagnaraRZH02},
a description of an NNC polyhedron $\cP \in \Pset_n$
in terms of generators can be obtained by also taking into account
its \emph{closure points}, i.e., points that belong to the topological
closure of the polyhedron, but are not necessarily included in the
polyhedron itself.
Namely, the results by Minkowski and Weil can be generalized
to the case of NNC polyhedra~\cite[Theorem~4.4]{BagnaraHZ05FAC}:
we can extend the generator system with a finite set $C$ of closure points,
obtaining $\cG = \langle L, R, C, P \rangle$ and $\cP = \gen(\cG)$, where
\[
  \gen\bigl( \langle L, R, C, P \rangle \bigr)
    \defeq
      \sset{
        L \vect{\lambda} + R \vect{\rho} + C \vect{\gamma} + P \vect{\pi}
          \in \Rset^n
      }{
       \vect{\lambda} \in \Rset^{\ell},
       \vect{\rho} \in \nonnegRset^r, \\
       \vect{\gamma} \in \nonnegRset^c,
       \vect{\pi} \in \nonnegRset^p,
       \vect{\pi} \neq \vect{0}, \\
       \textstyle{\sum_{i=1}^c \gamma_i + \sum_{i=1}^p \pi_i = 1}
     }.
\]

When needed for notational convenience,
we will split a constraint system into three components
$\cC = \langle \cC_{\mathord{=}}, \cC_{\mathord{\geq}}, \cC_{\mathord{>}} \rangle$;
even in this case, as done for the generators,
we will abuse the notation for set operator and relation symbols.

%% file: old-repr.tex
The DD method provides a solid theoretical base
for the representation and manipulation of topologically closed
convex polyhedra in $\CPset_n$. As mentioned in Section~\ref{sec:prelims},
at the implementation level the polyhedra are actually mapped into
polyhedral cones in $\CPset_{n+1}$ by homogenization, but it is not difficult
for software libraries to make this detail completely transparent
to the end user: in practice, the library developers have to add
some syntactic sugar to the input and output routines for constraints
and generators, also hiding the positivity constraint.

Things are less straightforward when considering the case
of NNC polyhedra.
To start with, many implementations of the DD method
do not support NNC polyhedra at all.
Also, to the best of our knowledge, the few supported
implementations of the domain of NNC polyhedra
based on the DD method
(that is, the NewPolka domain embedded in the Apron library
and the \texttt{NNC\_Polyhedron} domain in the Parma Polyhedra Library)
adopt an \emph{indirect} representation:
namely, each NNC polyhedron $\cP \in \Pset_n$
is mapped into a closed polyhedron $\cR \in \CPset_{n+1}$.
The mapping encodes the strict inequality constraints by means of
an additional space dimension (playing the role of a \emph{slack variable});
the new space dimension, usually denoted as $\epsilon$,
needs to be non-negative and bounded from above,%
\footnote{An alternative representation can be adopted where
the $\epsilon$ dimension is unbounded
from below~\cite{BagnaraHZ02a,BagnaraHZ05FAC}.}
i.e., the constraints $0 \leq \epsilon \leq 1$ are added
to the topologically closed representation $\cR$
(called $\epsilon$-representation)
of the NNC polyhedron $\cP$.

The inverse map
$\fund{\epssem{\cdot}}{\CPset_{n+1}}{\Pset_n}$
from an $\epsilon$-representation $\cR$ to
the represented NNC polyhedron $\cP$
is obtained by only considering
the points of $\cR$ having a strictly positive coordinate for
the $\epsilon$ dimension:
\[
\cP = \epssem{\cR}
  \defeq
    \bigl\{\,
      \vect{x} \in \Rset^n
    \bigm|
      (\vect{x}^\transpose, \epsilon)^\transpose \in \cR, \epsilon > 0
    \,\bigr\}.
\]

This encoding of NNC polyhedra into closed polyhedra was initially
proposed in~\cite{HalbwachsPR94,HalbwachsPR97} and later
reconsidered and studied in more detail
in~\cite{BagnaraHZ05FAC,BagnaraRZH02},
where a proper interpretation of the $\epsilon$ dimension for the
(extended) generator representation was provided.

\begin{figure}
\centering
{\scriptsize
\includegraphics{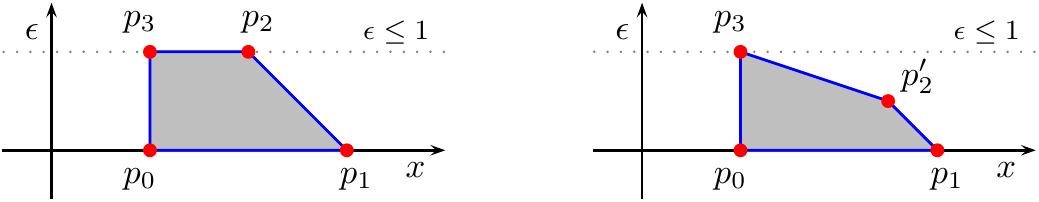}
}
\caption{Two $\epsilon$-representations in $\CPset_2$
for $\cP = \con(\cC) \in \Pset_1$,
where $\cC = \{ 1 \leq x, x < 3 \}$.}
\label{fig:eps-repr-examples}
\end{figure}

Besides showing its strengths,
the work in~\cite{BagnaraHZ05FAC,BagnaraRZH02} highlighted
the main weakness of the approach:
the DD pair in minimal form computed for an $\epsilon$-representation $\cR$,
when reinterpreted as encoding the NNC polyhedron $\cP = \epssem{\cR}$,
typically includes many redundant constraints and/or generators, leading to
a possibly high computational overhead.
To avoid this problem,
\emph{strong minimization procedures} were defined
in~\cite{BagnaraHZ05FAC,BagnaraRZH02}
that are able to detect and remove those redundancies;
in practice, these procedures map the representation $\cR$ into
a different representation $\cR'$ such that $\cP = \epssem{\cR'}$,
where $\cR'$ encodes no $\epsilon$-redundancies.

\begin{example}
Figure~\ref{fig:eps-repr-examples} shows two different
$\epsilon$-representations for the NNC polyhedron defined
by constraints $\cC = \{ 1 \leq x, x < 3 \}$.
In the $\epsilon$-representations,
the constraints having a zero coefficient for the slack variable $\epsilon$
encode a non-strict inequality, such as the one defining facet $[p_0,p_3]$,
corresponding to $1 \leq x$;
the constraints having a non-zero coefficient for $\epsilon$ encode
either the slack variable bounds $0 \leq \epsilon \leq 1$
or the proper strict inequalities,
such as the one defining facet $[p_1,p_2]$
(resp., $[p_1,p'_2]$ on the right hand side $\epsilon$-representation),
corresponding to $x < 3$.
Note that the facet $[p'_2, p_3]$
in the right hand side $\epsilon$-representation
is an example of $\epsilon$-redundant constraint,
since it is encoding the redundant strict inequality $x < 4$.
\end{example}

When carefully applying strong minimization procedures,
most of the overhead of the $\epsilon$-representation is thus avoided,
leading to implementations that easily meet the efficiency requirements
of many application contexts.%
\footnote{After being initially implemented and tested in the
Parma Polyhedra Library, these strong minimization procedures
have also been adopted in the Apron library.}
For the users of the libraries, the addition of the $\epsilon$ dimension
is almost unnoticed, to the point that quite often the domain of NNC
polyhedra is adopted even when not really needed (i.e., when a domain
of topologically closed polyhedra would be enough).
However, the approach described above still suffers from a few issues.
\begin{enumerate}
\item
At the implementation level, more work is needed to make the $\epsilon$
dimension \emph{transparent} to the end user and, as a matter of fact,
its adoption can sometimes become evident.
For instance, a strict constraint such as $x > 30$ may be encoded
as $2x - \epsilon \geq 60$, which is then shown to the user as the
(unsimplified) strict constraint $2x > 60$.%
\footnote{See \url{https://www.cs.unipr.it/mantis/view.php?id=428}.}
Besides being annoying, the growth in the magnitude of
the integer coefficients may cause a computational overhead.
\item
The $\epsilon$-representation brings with itself an \emph{intrinsic}
overhead: in any generator system for an $\epsilon$-polyhedron,
most of the ``proper'' points (those having a positive $\epsilon$ coordinate)
need to be paired with the corresponding ``closure'' point
(having a zero $\epsilon$ coordinate);
this systematically leads to almost doubling the size
of the generator system.
\item
The strong minimization procedures, even though effective,
interfere with the \emph{incremental} approach
of the DD conversion procedures.
After applying the strong minimization procedure on the constraint
(resp., generator) representation of a DD pair,
the dual generator (resp., constraint) representation is lost and,
in order to recover it,
the non-incremental conversion procedure needs to be applied once again.
This also implies that the strong minimization procedures
can not be fully integrated into the DD conversion procedures:
they are applied \emph{after} the conversions.
As a consequence, during the iterations of the conversion procedure,
the redundancies caused by the $\epsilon$-representation are not removed,
causing the computation of bigger intermediate results.
For the reasons above, the strong minimization procedures are not
systematically used in the implementation of the Parma Polyhedra Library;
rather, they are applied only when strictly needed for correctness.
Therefore, the end user is left with the responsibility of \emph{guessing}
whether or not the strong minimization procedures are
going to improve efficiency.
\end{enumerate}

The most important of the issues listed above were known
since~\cite{BagnaraRZH02}.
As a matter of fact, both~\cite{BagnaraHZ05FAC} and~\cite{BagnaraHZ09TCS}
put forward the possibility of devising an alternative approach
regarding the representation and manipulation of NNC polyhedra
in the DD framework.
Quoting from~\cite{BagnaraHZ09TCS}:
\begin{quote}
It would be interesting,
from both a theoretical and practical point of view,
to provide a more direct encoding of NNC polyhedra,
i.e., one that is not based on the use of slack variables [\dots]
\end{quote}
The main obstacle on the road towards such a goal is
the definition of a conversion procedure that is not only correct,
but also competitive with respect to the highly tuned implementations
available in software libraries such as Apron and
the Parma Polyhedra Library.
It is worth stressing that several experimental evaluations,
including recent ones~\cite{AssarfGHJLPR17},
confirm that the Parma Polyhedra Library is a state-of-the-art
implementation of the DD method for a wide spectrum of
application contexts.

%% file: new-rep.tex
As briefly recalled in Section~\ref{sec:prelims},
an NNC polyhedron can be described by using an extended constraint system
$\cC = \langle \cC_{\mathord{=}}, \cC_{\mathord{\geq}}, \cC_{\mathord{>}} \rangle$,
possibly containing strict inequalities,
and/or an extended generator system
$\cG = \langle L, R, C, P \rangle$,
possibly containing closure points.
These representations are said to be \emph{geometric},
meaning that they provide a precise description of
the position of all the elements in the constraint/generator system.

For a closed polyhedron $\cP \in \CPset_n$,
the use of completely geometric representations is an adequate choice:
it is possible to provide a DD pair $(\cC, \cG)$ that is ``canonical''.%
\footnote{Strictly speaking, the canonical form for constraints
(resp., generators) still depends on the specific representation
chosen for the non-redundant set of equality constraints
(resp., generating lines). Even those can be made canonical and
each software library typically provides its own canonical form.}
In the case of an NNC polyhedron $\cP \in \Pset_n$,
the adoption of a completely geometric representation can be seen
as an overkill, since the knowledge of the precise geometric position
of some of the elements is not really needed.

\begin{figure}
\centering
{\scriptsize
\includegraphics{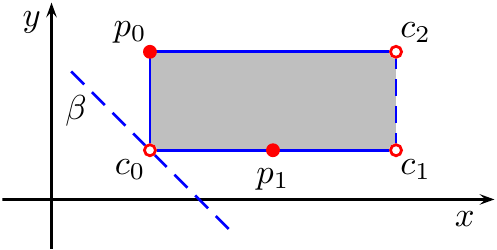}
}
\caption{An NNC polyhedron having no ``canonical'' geometric representations.}
\label{fig:nnc-non-canonical}
\end{figure}

\begin{example}
Consider the NNC polyhedron $\cP \in \Pset_2$
in Figure~\ref{fig:nnc-non-canonical}, where
the (strict) inequality constraints are denoted by (dashed) lines and
the (closure) points are denoted by (unfilled) circles.
The polyhedron can be seen to be described by generator system
$\cG = \langle L, R, C, P \rangle$, where
$L = R = \emptyset$, $C = \{ c_0, c_1, c_2 \}$ and $P = \{ p_0, p_1 \}$.
However, there is no need to know the precise position of point $p_1$,
since it can be replaced by any other point on the open segment $(c_0, c_1)$.
Similarly, when considering the constraint representation,
there is no need to know the exact slope of the strict inequality
constraint $\beta$, as it can be replaced by any other
strict inequality that is satisfied by all the points in $\cP$ and
saturated by closure point $c_0$.
\end{example}

In other words, some of the elements in the geometric representations
of NNC polyhedra are better described by combinatorial information,
rather than geometric. The following section introduces
the terminology and notation needed to reason on this combinatorial
information.

\subsection{The combinatorial structure of convex polyhedra}
\label{subsec:face-lattice}

A linear inequality or equality constraint
$\beta = (\vect{a}^\transpose \vect{x} \relop b)$
is said to be \emph{valid} for the polyhedron $\cP \in \CPset_n$
if all the points in $\cP$ satisfy $\beta$;
for each such $\beta$, the subset
\(
  F = \{\, \vect{p} \in \cP \mid \vect{a}^\transpose \vect{p} = b \,\}
\)
is a \emph{face} of $\cP$.
We write $\CFaces_{\cP}$, omitting the subscript when clear from context,
to denote the finite set of faces of $\cP \in \CPset_n$;
the set $\CFaces_\cP$ is a sublattice of $\CPset_n$,
having the empty face as bottom element and the whole polyhedron $\cP$
as top element.
Note that we have
\[
  \cP = \bigunion \bigl\{\, \relint(F) \bigm| F \in \CFaces_\cP \,\bigr\}.
\]
The face lattice is also known as
the combinatorial structure of the polyhedron.
If the polyhedron is bounded (i.e., it is a polytope, having
no rays and lines), then the lattice is atomistic,
meaning that each face can be obtained as the convex polyhedral hull
of the vertices contained in the face.

Even in the case of an NNC polyhedron $\cP \in \Pset_n$
it is possible to define the finite set $\NNCFaces_\cP$
of its faces, which is a sublattice of $\Pset_n$;
hence, each face is an NNC polyhedron and, as before, we have
\[
  \cP = \bigunion \bigl\{\, \relint(F) \bigm| F \in \NNCFaces_\cP \,\bigr\}.
\]
In this case, however, the lattice may be non-atomistic
even when the polyhedron is bounded.
Letting $\cQ = \cl(\cP)$,
the closure operator
$\fund{\cl}{\NNCFaces_\cP}{\CFaces_{\cQ}}$
maps each NNC face of $\cP$ into a distinct,
corresponding (closed) face of $\cQ$.
The image $\cl(\NNCFaces_\cP)$ is a join sublattice of $\CFaces_\cQ$;
meets are generally not preserved, since there may exist
$F_1, F_2 \in \NNCFaces_\cP$ such that
\[
  F_1 \inters F_2 = \emptyset \neq \cl(F_1) \inters \cl(F_2).
\]
The image of the set of non-empty faces
$\cl\bigl(\NNCFaces_\cP \setminus \{ \emptyset \} \bigr)$
is an upward closed subset of $\CFaces_\cQ$;
hence, it can be efficiently described by recording
just the set of its minimal elements.
For each NNC face $F \sseq \cP$ corresponding to one
of these minimal elements
(that is, for each atom of the $\NNCFaces_\cP$ lattice),
we have $F = \relint(F)$.
As a consequence, the combinatorial structure of $\cP \in \Pset_n$
can be described by integrating the combinatorial structure
of its topological closure $\cQ \in \CPset_n$
with the information identifying the atoms of $\NNCFaces_\cP$.

\begin{example}
Consider the polyhedron $\cP$ in Figure~\ref{fig:nnc-non-canonical}.
The lattice $\NNCFaces_\cP$ has two atoms:
the 0-dimension face $\{ p_0 \}$
and the 1-dimension open segment $(c_0, c_1)$;
note that both atoms are relatively open sets.
Also note that, even if $\cP$ is an NNC polytope,
the lattice is not atomistic: for instance,
the half-open segment $(c_0, p_0]$ is a 1-dimension face
that can not be obtained by joining the atoms.
\end{example}

\subsection{Skeleton and non-skeleton of an NNC polyhedron}

Let $\cP \in \Pset_n$ be an NNC polyhedron and $\cQ = \cl(\cP) \in \CPset_n$
be its topological closure.
As explained above, a description of $\cP$ can be obtained
by combining a geometric representation of $\cQ$,
which will be called the \emph{skeleton}%
\footnote{This term is unrelated to the concept of $p$-skeleton
used in algebraic topology.}
component,
with some combinatorial information related to $\NNCFaces_\cP$
(the \emph{non-skeleton} component).
We now provide formal definitions that allow for splitting a
fully geometric representation for $\cP$ into these two components.
For exposition purposes, here we will consider the generator system
representation only;
the definitions for the constraint system representation
are similar and will be briefly described in a later section.

\begin{definition}[Skeleton of a generator system]
\label{def:skeleton}
Let $\cG = \langle L, R, C, P \rangle$ be a generator system
in minimal form, $\cP = \gen(\cG)$ and $\cQ = \cl(\cP)$.
The \emph{skeleton} of $\cG$ is the generator system
\[
  \SK_\cQ = \skel(\cG) \defeq \langle L, R, C \union \SP, \emptyset \rangle,
\]
where $\SP \sseq P$ is the set of points that can not be obtained
as a combination of the other generators in $\cG$.
\end{definition}

Note that the skeleton has no points at all,
so that $\gen(\SK_\cQ) = \emptyset$.
However, we can define a variant function `$\mathord{\fullgen}$',
that reinterprets the closure points to be points,
\[
  \fullgen\bigl( \langle L, R, C, P \rangle \bigr)
    \defeq
      \gen\bigl( \langle L, R, \emptyset, C \union P \rangle \bigr),
\]
so as to obtain the following result.

\begin{proposition}
\label{prop:skeleton}
Let $\cP = \gen(\cG)$ and $\cQ = \cl(\cP)$. Then
\[
  \fullgen(\cG) = \fullgen(\SK_\cQ) = \cQ.
\]
Also, there does not exist $\cG' \subset \SK_\cQ$
such that $\fullgen(\cG') = \cQ$.
\end{proposition}

In other words, the skeleton of an NNC polyhedron can be seen
to provide a non-redundant representation of its topological closure.
The elements of $\SP \sseq P$ are called \emph{skeleton points};
the non-skeleton points in $P \setminus \SP$ are redundant when
representing the topological closure, since they can be obtained
by combining the lines in $L$, the rays in $R$ and the closure points in $C$;
these \emph{non-skeleton points} are the elements in $\cG$
that need not be represented geometrically.

\begin{example}
For the polyhedron in Figure~\ref{fig:nnc-non-canonical},
\(
  \SK_\cQ
    = \bigl\langle
        \emptyset, \emptyset, \{ c_0, c_1, c_2, p_0 \}, \emptyset
      \bigr\rangle
\),
so that $p_0$ is a skeleton point and $p_1$ is a non-skeleton point
(it can be generated by combining $c_0$ and $c_1$).
\end{example}

Having modeled the skeleton component for
$\cP = \gen\bigl( \langle L, R, C, P \rangle \bigr)$,
we now turn our attention to the non-skeleton component.
As discussed in Section~\ref{subsec:face-lattice},
our goal is to provide a combinatorial representation for
the set of points $P$.
Reasoning slightly more generally,
consider a point $\vect{p} \in \cQ = \cl(\cP)$
(not necessarily in $P$).
There exists a single face $F \in \CFaces_{\cQ}$
such that $\vect{p} \in \relint(F)$.
By definition of function `$\gen$',
point $\vect{p}$ behaves as a \emph{filler} for $\relint(F)$,
meaning that, when combined with the skeleton,
it generates $\relint(F)$.
Note that $\vect{p}$ also behaves as a filler
for the relative interiors of all the faces in the set $\upcl F$.
The choice of $\vect{p} \in \relint(F)$ is actually arbitrary:
any other point of $\relint(F)$ would be equivalent as a filler.

\begin{proposition}
\label{prop:multiple-mater}
Consider a polyhedron $\cP = \gen(\cG)$,
where $\cG = \langle L, R, C, P \rangle$.
For $\vect{p} \in P$,
let $F$ be the face of $\cQ = \cl(\cP)$
such that $\vect{p} \in \relint(F)$;
let $\vect{p}' \in \relint(F)$,
and $P' = P \setminus \{ \vect{p} \} \union \{ \vect{p}' \}$.
Then $\cP = \gen\bigl( \langle L, R, C, P' \rangle \bigr)$.
\end{proposition}

A less arbitrary representation for $\relint(F)$ is thus provided
by its own skeleton $\SK_F \sseq \SK_{\cQ}$;
namely, each (geometric) filler $\vect{p} \in \cP$
can be mapped into a more abstract (combinatorial) representation,
the subset of $\SK_{\cQ}$ identifying the corresponding face.
For each face $F \in \CFaces_{\cQ}$, we say that
the skeleton subset $\SK_F \sseq \SK_{\cQ}$
is the \emph{support} for the points in $\relint(F)$ and that
any point $\vect{p}' \in \relint\bigl(\fullgen(\SK_F)\bigr) = \relint(F)$
is a \emph{materialization} of $\SK_F$.

\begin{definition}[Support sets for a skeleton]
Let $\SK$ be the skeleton of an NNC polyhedron and
let $\cQ = \fullgen(\SK) \in \CPset_n$.
Then the set $\NSset_{\SK}$ of all supports for $\SK$ is defined as
\[
  \NSset_{\SK}
    \defeq
      \{\,
        \SK_F \sseq \SK
      \mid
        F \in \CFaces_{\cQ}
      \,\}.
\]
\end{definition}
By definition, the set $\NSset_{\SK}$
is a lattice isomorphic to $\CFaces_\cQ$;
we will drop the subscripts $\SK$ and $\cQ$ when clear from context.

We now define a pair of abstraction and concretization functions
mapping a subset of the (geometric) points of an NNC polyhedron
into the set of supports that are filled by these points,
and vice versa.

\begin{definition}[Filled supports]
\label{def:alpha-gamma}
Let $\SK$ be the skeleton of the polyhedron $\cP \in \Pset_n$,
$\cQ = \cl(\cP)$ and $\NSset$ be the corresponding set of supports.
The abstraction function
$\fund{\alpha_{\SK}}{\wp(\cQ)}{\wpup(\NSset)}$
is defined, for each $S \sseq \cQ$, as
\[
  \alpha_{\SK}(S)
    \defeq
      \bigcup
        \bigl\{\,
          \upcl \SK_F
        \bigm|
          \exists \vect{p} \in S, F \in \CFaces \st \vect{p} \in \relint(F)
        \,\bigr\}.
\]
The concretization function
$\fund{\gamma_{\SK}}{\wpup(\NSset)}{\wp(\cQ)}$,
for each $\NS \in \wpup(\NSset)$, is defined as
\[
  \gamma_{\SK}(\NS)
    \defeq
      \bigcup
        \Bigl\{\,
          \relint\bigl(\fullgen(\ns)\bigr)
        \Bigm|
          \ns \in \NS
        \,\Bigr\}.
\]
\end{definition}

\begin{proposition}
\label{prop:Galois-connection}
The pair of functions $(\alpha_{\SK}, \gamma_{\SK})$
is a Galois connection.
\end{proposition}

By Proposition~\ref{prop:Galois-connection},
the composition $(\gamma_\SK \comp \alpha_\SK)$
is an upper closure operator mapping
each non-empty set of points $S \sseq \cQ$
into the smallest NNC polyhedron containing $S$
and having $\SK$ as the skeleton component.
In particular, the following result holds.

\begin{proposition}
\label{prop:fixpoint}
Let $\cP = \gen\bigl( \langle L, R, C, P \rangle \bigr) \in \Pset_n$
and let $\SK$ be the corresponding skeleton component.
Then $\cP = (\gamma_\SK \comp \alpha_\SK)(P)$.
\end{proposition}

The non-skeleton component of a geometrical generator system,
can be abstracted by `$\alpha_\SK$' and described
as a combination of skeleton generators.
\begin{definition}[Non-skeleton of a generator system]
\label{def:non-skel}
Let $\cP \in \Pset_n$ be defined by generator system
$\cG = \langle L, R, C, P \rangle$
and let $\SK$ be the corresponding skeleton component.
The \emph{non-skeleton} component of $\cG$ is defined as
\(
  \NS_\cG \defeq \alpha_\SK(P)
\).
\end{definition}
Even in this case, we will drop the subscript when clear from context.
Note that, by definition of the abstraction function `$\alpha_\SK$',
the non-skeleton component $\NS$ contains an upward closed set of supports,
therefore representing \emph{all} the faces of the NNC polyhedron.

\begin{example}
We now show the non-skeleton component for the polyhedron
in Figure~\ref{fig:nnc-non-canonical}. Since in this polyhedron we
have no rays and no lines, we will adopt a simplified notation,
identifying each support with the set of its closure points.
By Definition~\ref{def:alpha-gamma}, we have:
\begin{align*}
  \alpha_\SK\bigl( \{ p_0 \} \bigr)
    &= \bigl\{\,
         \{ p_0 \}, \{ c_0, p_0 \}, \{ c_2, p_0 \}, \{ c_0, c_1, c_2, p_0 \}
       \,\bigl\}, \\
  \alpha_\SK\bigl( \{ p_1 \} \bigr)
    &= \bigl\{\,
         \{ c_0, c_1 \}, \{ c_0, c_1, c_2, p_0 \}
       \,\bigl\};
\end{align*}
hence, the non-skeleton component is computed as
\begin{align*}
  \NS_\cG
    &= \alpha_\SK\bigl( \{ p_0, p_1 \} \bigr)
     = \bigl\{\,
         \{ p_0 \}, \{ c_0, p_0 \}, \{ c_2, p_0 \}, \{ c_0, c_1, c_2, p_0 \},
         \{ c_0, c_1 \}
       \,\bigr\}.
\end{align*}
The minimal elements in $\NS_\cG$ are the supports
$\{ p_0 \}$ and $\{ c_0, c_1 \}$,
which can be seen to describe the atoms of the face lattice $\NNCFaces_\cP$.
\end{example}

By combining Definition~\ref{def:non-skel}
with Proposition~\ref{prop:fixpoint}
we obtain the following result,
stating that the new representation
is semantically equivalent to the fully geometric one.
\begin{corollary}
For a polyhedron $\cP = \gen(\cG) \in \Pset_n$,
let $\langle \SK, \NS \rangle$
be the skeleton and non-skeleton components for $\cG$.
Then $\cP = \gamma_\SK(\NS)$.
\end{corollary}

%% file: conv.tex
When working with direct representations of NNC polyhedra,
the Chernikova's conversion algorithm needs to be extended
to properly handle closure points and strict inequalities.

A first attempt in this direction was developed in~\cite{Perri12th}.
In that case, the $\epsilon$-less encoding for constraints and generators
was not distinguishing the skeleton and non-skeleton components,
thereby adopting geometric-only representations.
The main difference with respect to the classical conversion algorithm
for closed polyhedra was in the combination phase,
where the sets of generators $\cG^+$, $\cG^-$ are processed
to produce the new set of generators $\cG^{\star}$:
this phase was extended in~\cite{Perri12th}
to perform a systematic case analysis
on the generator kinds and to also consider the set $\cG^0$.
Even though the resulting algorithm is correctly specified,
it suffers from a high computational overhead because,
as highlighted in~\cite{Zaffanella17},
the new combination phase needs to also consider pairs of generators
that are \emph{not} adjacent; this prevents the adoption
of the key optimizations that were developed for the closed polyhedra case,
making the overall approach infeasible from a practical point of view.

The new representation described in Section~\ref{sec:new-repr},
by distinguishing the skeleton and non-skeleton components,
allows for a corresponding separation in the conversion procedure:
while the skeleton component can be handled following the classical
combination procedure for closed polyhedra,
the non-skeleton will be managed using a few brand new procedures
that can correctly deal with closure points and strict
inequalities without incurring into a significant overhead.

As already pointed out in Section~\ref{sec:new-repr},
we will focus on the conversion from constraints to generators.
The conversion working the other way round will be obtained,
as usual, by applying duality arguments.

The {\sc conversion} function is shown as Pseudocode~\ref{alg:conversion}.
In the following, we will describe its main steps,
first introducing some implementation details
and then explaining the auxiliary functions and procedures.

\subsection{Encoding the new representation}
\label{sec:conv-new-repr-encoding}

In Section~\ref{sec:new-repr} it was shown how
the geometric generator system $\cG$ can be equivalently represented
by the pair $\langle \SK, \NS \rangle$,
where $\SK = \langle  L, R, C \union \SP, \emptyset \rangle$
is the skeleton component
and $\NS \sseq \wpup(\NSset)$ is the non-skeleton component.
We now discuss a few minor adaptations to this representation that
are meant to result in efficiency improvements at the implementation level.

First, observe that every support $\ns \in \NS$
always includes all of the lines in the $L$ skeleton component;
hence, these lines can be left \emph{implicit} in the representation
of the supports in $\NS$.
Note that, even after removing the lines,
each $\ns \in \NS$ is still a non-empty set,
since it includes at least one closure point.

When lines are implicit, those supports $\ns \in \NS$
that happen to be singletons%
\footnote{Since the support $\ns$ is a subset of the skeleton $\SK$,
by `singleton' here we mean a system
\(
  \ns = \bigl\langle
          \emptyset, \emptyset, \{ \vect{p} \}, \emptyset
        \bigr\rangle
\).}
can be seen to play a special role:
they correspond to the combinatorial encoding of
the skeleton points in $\SP$ (see Definition~\ref{def:skeleton}).
These points are not going to benefit from the combinatorial
representation, since their geometric position is uniquely identified
(modulo the lines component).
Therefore, we will remove them from the non-skeleton $\NS$
and directly include them in the point component of the skeleton $\SK$;
namely, the skeleton
$\SK = \langle L, R, C \union \SP, \emptyset \rangle$
will be actually represented as
$\SK = \langle L, R, C, \SP \rangle$.
We stress that this is only done as an optimization:
the formalization presented in Section~\ref{sec:new-repr} is still
valid, with just a minor adaptation to the definition of the
function `$\gamma$', which is replaced by the following:
\[
  \gamma'_{\SK}(\NS) \defeq \gen(\SK) \cup \gamma_{\SK}(\NS).
\]

We also remark that, at the implementation level,
each support $\ns \in \NS$ can be encoded by using a \emph{set of indices}
on the data structure representing the skeleton component $\SK$.
Since $\NS$ is a finite upward closed set,
the representation only needs to record its minimal elements.
In this low level representation, the non-minimal elements
can be efficiently identified (and removed) by performing
appropriate inclusion tests on these sets of indices.
When also considering the optimization for skeleton points
mentioned before, we can adopt the following definition of redundancy.

\begin{definition}[Redundant support]
A support $\ns \in \NS$ is said to be
\emph{redundant in $\langle \SK, \NS \rangle$}
if there exists $\ns' \in \NS$ such that $\ns' \subset \ns$
or if $\ns \inters \SP \neq \emptyset$,
where $\SK = \langle L, R, C, \SP \rangle$.
\end{definition}
In the following, we will write $\NS_1 \oplus \NS_2$ to denote
the non-redundant union of the support sets $\NS_1, \NS_2 \sseq \NSset_\SK$.

\input{pseudo-conv}

\subsection{Processing the skeleton component}

From a high level point of view,
the {\sc conversion} function in Pseudocode~\ref{alg:conversion}
follows the same structure as the classical conversion procedure
for closed polyhedra:
it incrementally processes each of the input constraints
$\beta \in \cC_{\mathit{in}}$
keeping the generator system $\langle \SK, \NS \rangle$ up-to-date.
In this section, we focus on the handling
of the skeleton component $\SK = \langle L, R, C, \SP \rangle$.

The first processing step (\cref{conv.sk-part}) of the main loop
is the partitioning of the skeleton $\SK$
according to the signs of the scalar products with constraint $\beta$.
Since the skeleton component is entirely geometric,
it can be split into $\SK^+$, $\SK^0$ and $\SK^-$
exactly as done in the Chernikova's algorithm.
In the pseudocode, this partition info is kept implicit
inside the data structure encoding $\SK$: we will freely use
the superscripts to refer to each component when needed.

Note that \crefrange{conv.line-b}{conv.line-e}
of the {\sc conversion} function
are meant to take care of a line violating $\beta$,
whereas \crefrange{conv.spec-b}{conv.spec-e}
are meant to efficiently handle
those special cases when $\SK^+$ or $\SK^-$ happens to be empty;
these will be briefly discussed later on.
Hence, the second main processing step for the skeleton component
occurs in \crefrange{conv.sk-comb-b}{conv.sk-comb-e},
where the generators in $\SK^+$ and $\SK^-$
are combined to produce $\SK^\star$, which is then merged into $\SK^0$.
This step too is quite similar to the one for closed polyhedra
described in Section~\ref{sec:prelims},
except that we now have to consider how the different generator kinds
combine with each other, according to the kind of constraint $\beta$:
the systematic case analysis is presented in Table~\ref{tab:sk-combine-kind}.
The table shows that, for instance, when processing a non-strict
inequality $\beta_\geq$, if we combine a closure point in $\SK^+$
with a ray in $\SK^-$  we shall obtain a closure point in $\SK^\star$
(row 3, column 6).

\begin{table}
\centering
\begin{tabular}{c|c|ccccccccc}
  & $\SK^+$
  & R & R & R & C & C & C & SP & SP & SP \\
  & $\SK^-$
  & R & C & SP & R & C & SP & R & C & SP \\
\hline
$\beta_=$ or $\beta_\geq$
  & \multirow{2}{*}{$\SK^\star$}
  & R & C & SP & C & C & SP & SP & SP & SP \\
 $\beta_>$
 &
 & R & C & C & C & C & C & C & C & C \\
 \multicolumn{11}{c}{}
\end{tabular}
\caption{Case analysis for function `$\combine{\beta}$'
when adding an equality ($\beta_=$), a non-strict ($\beta_\geq$) or a
strict ($\beta_>$) inequality constraint
to a pair of generators from $\SK^+$ and $\SK^-$
(R = ray, C = closure point, SP = skeleton point).}
\label{tab:sk-combine-kind}
\end{table}

A crucial observation regarding this combination phase is that,
since it is restricted to work on the skeleton component only,
it can safely apply the adjacency tests to quickly get rid
of all those combinations that would introduce redundant elements
(for the skeleton component).
Also note how the direct inclusion of the skeleton points $\SP$ in $\SK$
(as discussed in Section~\ref{sec:conv-new-repr-encoding}),
besides simplifying the non-skeleton representation,
allows for processing them using the adjacency tests.
Nonetheless, since the points in $\SP$ should behave as fillers,
they will have to be properly reconsidered when processing
the non-skeleton component $\NS$.

The final processing steps for the skeleton component,
occurring in \crefrange{conv.final-b}{conv.final-e},
are those meant to update the generator system for the next iteration.
The new skeleton is computed according to the constraint kind,
similarly to what done in the closed polyhedra case.
However, an additional processing step (\cref{conv.p-to-cp})
is needed for the case of a strict inequality constraint:
the helper function
\[
  \pbecomecp\bigl(\langle L, R, C, \SP \rangle\bigr)
    \defeq \langle L, R, C \union \SP, \emptyset \rangle
\]
applied to $\SK^0$,
makes sure that all of the skeleton points saturating $\beta$
are transformed into closure points having the same position.

\subsection{Processing the non-skeleton component}

We now consider the handling of the non-skeleton component $\NS$,
which is clearly where the new algorithm significantly differs from the
corresponding algorithm for closed polyhedra.

\subsubsection*{Partitioning.}
The first processing step (line 4) is the partitioning of the
supports in $\NS$, so as to detect their position with respect
to the constraint $\beta$.
To this end, we can exploit the partition info already
computed for the skeleton $\SK$ to obtain the corresponding
partition info for $\NS$, without computing any additional scalar product.
Namely, each support $\ns \in \NS$ is classified as follows:
\begin{align*}
  \ns \in \NS^+
    &\Longleftrightarrow
      \ns \subseteq (\SK^+ \union \SK^0)
        \land
      \ns \inters \SK^+ \neq \emptyset; \\
  \ns \in \NS^0
    &\Longleftrightarrow
      \ns \subseteq \SK^0; \\
  \ns \in \NS^-
    &\Longleftrightarrow
      \ns \subseteq (\SK^- \union \SK^0)
        \land
      \ns \inters \SK^- \neq \emptyset; \\
  \ns \in \NS^\pm
    &\Longleftrightarrow
      \ns \inters \SK^+ \neq \emptyset
        \land
      \ns \inters \SK^- \neq \emptyset.
\end{align*}
Note that the partitioning above is fully consistent
with respect to the one computed for skeleton elements.
For instance, if $\ns \in \NS^+$, then for every possible
materialization $\vect{p} \in \relint(\fullgen(\ns))$
the scalar product of $\vect{p}$ and $\beta$ is strictly positive.
Things are similar when $\ns \in \NS^0$ and $\ns \in \NS^-$.
The supports in $\NS^\pm$ are those whose materializations can
indifferently satisfy, saturate or violate the constraint $\beta$
(i.e., the corresponding face \emph{crosses} the constraint hyperplane).
As did for the skeleton, even in this case the partition info
is kept implicit inside the data structure encoding $\NS$.

\input{pseudo-promote}

As said before, we delay for the moment the discussion
of \crefrange{conv.line-b}{conv.spec-e} of the {\sc conversion} function,
proceeding directly to explain \cref{conv.move-ns,conv.create-ns},
where we find the calls of the two main functions
processing the non-skeleton component.
A set $\NS^\star$ of brand new supports is built
as the union of the contributes provided by
functions {\sc move-ns} and {\sc create-ns}.
This set, which contains the supports generated in a given iteration
of the main loop, will be later merged into the appropriate portions
of the non-skeleton component, chosen according to the constraint kind
(see \crefrange{conv.final-b}{conv.final-e}).
The final processing step of the main loop (\cref{conv.promote-sing})
calls helper procedure {\sc promote-singletons}
(shown in Pseudocode~\ref{alg:promote}),
making sure that all singleton supports
get promoted to skeleton points.

\subsubsection*{Moving supports.}

The {\sc move-ns} function, shown in Pseudocode~\ref{alg:move},
processes the supports in $\NS^\pm$.
As hinted by its name, the goal of this function is to ``move''
the fillers of the faces that are crossed by the new constraint,
making sure they lie on the correct side.

\input{pseudo-move}

Let $\ns \in \NS^\pm$ and consider the face $F = \relint(\fullgen(\ns))$.
Note that $F$ is a face of the polyhedron \emph{before}
the addition of the new constraint $\beta$;
at this point, the elements in $\SK^\star$ have been added to $\SK^0$,
but this change still has to be propagated
to the non-skeleton component $\NS$.
Therefore, we compute the \emph{support closure} `$\suppcl_{\SK}(\ns)$'
of the support $\ns$ according to the updated skeleton $\SK$.
Intuitively, $\suppcl_{\SK}(\ns) \sseq \SK$ is the subset
of all the skeleton elements that are included in face $F$.

At the implementation level, the support closure operator
can be efficiently computed by exploiting the same
\emph{saturation information} that is needed to quickly perform
the adjacency tests.
Namely, given the constraints $\cC$ and the generators $\cG$,
we can define the functions
\begin{align*}
  \satinter_{\cC}(\cG)
    &\defeq
      \{\,
        \beta' \in \cC
      \mid
        \forall g \in \cG \itc \text{$g$ saturates $\beta'$}
      \,\}, \\
  \satinter_\cG(\cC)
    &\defeq
      \{\,
        g \in \cG
      \mid
        \forall \beta' \in \cC \itc \text{$g$ saturates $\beta'$}
      \,\}.
\end{align*}
Then, if $\cC$ and $\SK = \langle L, R, C, \SP \rangle$
are the constraint system and the skeleton generator system
defining the polyhedron, for each $\ns \in \NS$ we can compute
the support closure as follows~\cite{KaibelP02}:
\begin{align*}
  \suppcl_\SK(\ns)
    &\defeq
      \satinter_\SK\bigl(\satinter_{\cC}(\ns)\bigr) \setdiff L.
\end{align*}

Face $F$ is intuitively split by constraint $\beta$
into the three subsets $F^+$, $F^0$ and $F^-$.
When $\beta$ is a strict inequality, only $\cF^+$ shall
be kept in the polyhedron;
when the new constraint is a non-strict inequality,
both $\cF^+$ and $\cF^0$ shall be kept.
When working with the updated support,
a non-skeleton representation for these subsets can be obtained by
\emph{projecting} the support on the corresponding portions
of the skeleton.
Namely, we can define the function
\[
  \proj^\beta_\SK(\ns)
    \defeq
      \begin{cases}
        \ns \setdiff \SK^-, & \text{if $\beta$ is a strict inequality;} \\
        \ns \inters \SK^0,  & \text{otherwise.}
      \end{cases}
\]
Since the projection operator is applied \emph{after} having
computed the support closure, when $\beta$ is a non-strict inequality
we have $\ns \inters \SK^0 \neq \emptyset$;
hence, the support of $F^0$ is a subset of the support of $F^+$
and $\proj^\beta_\SK(\ns)$ will be a filler for $F^+$ too.

To summarize, by composing support closure and projection
in~\cref{move.proj-cl} of {\sc move-ns},
each support in $\NS^\pm$ is moved to the correct side of $\beta$.

\begin{figure}
\centering
{\scriptsize
\includegraphics{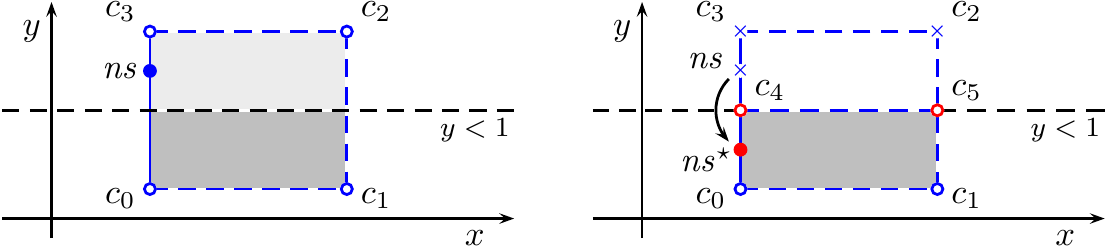}
}
\caption{Application of {\sc move-ns} to $\ns \in \NS^\pm$
when adding a strict inequality.}
\label{fig:s-move}
\end{figure}

\begin{example}
\label{ex:s-move}
Consider the polyhedron $\cP \in \Pset_2$
in the left hand side of Figure~\ref{fig:s-move},
described by the skeleton and non-skeleton components
$\langle \SK, \NS \rangle$.
The skeleton $\SK = \langle \emptyset, \emptyset, C, \emptyset \rangle$
is composed by the four closure points in $C = \{ c_0, c_1, c_2, c_3 \}$;
the non-skeleton $\NS = \{ \ns \}$ contains a single support
$\ns = \{ c_0, c_3 \}$, which makes sure that
the open segment $(c_0, c_3)$ is included in $\cP$;
in the figure, we show just one of the many possible
materializations for $\ns$.

When processing the strict inequality constraint $\beta = (y < 1)$,
we obtain the polyhedron in the right hand side of the figure.
In the skeleton phase of the {\sc conversion} function
the adjacent skeleton generators are combined:
$c_4$ (combining $c_0 \in \SK^+$ and $c_3 \in \SK^-$)
and $c_5$ (combining $c_1 \in \SK^+$ and $c_2 \in \SK^-$)
are added to $\SK^0$.
Since the non-skeleton support $\ns$ belongs to $\NS^\pm$,
it is processed in the {\sc move-ns} function:
\begin{align*}
  \ns^*
    &= \proj^\beta_\SK\bigl( \suppcl_\SK(\ns) \bigr) \\
    &= \suppcl_\SK\bigl( \{ c_0, c_3 \} \bigr) \setdiff \SK^- \\
    &= \{ c_0, c_3, c_4 \} \setdiff \{ c_2, c_3 \} \\
    &= \{ c_0, c_4 \}.
\end{align*}
Intuitively, we have moved $\ns$ to $\ns^\star$:
again, for the new support we show only one of its many possible
materializations, but it is clear that now they all
satisfy constraint $\beta$.
\end{example}

\begin{figure}
\centering
{\scriptsize
\includegraphics{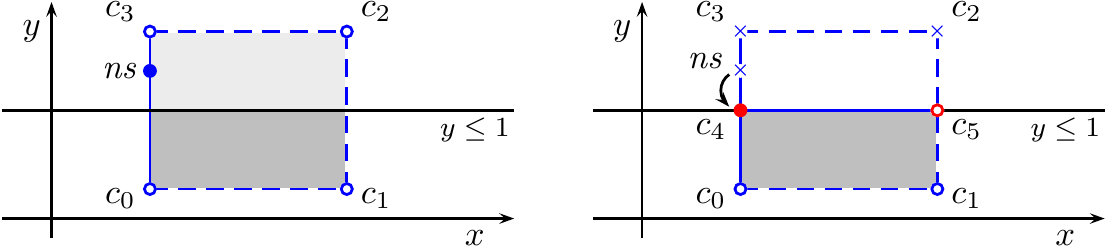}
}
\caption{Application of {\sc move-ns} to $\ns \in \NS^\pm$,
adding a non-strict inequality.}
\label{fig:ns-move}
\end{figure}
\begin{example}
In the left hand side of Figure~\ref{fig:ns-move},
we reconsider the same polyhedron of Example~\ref{ex:s-move},
but we now add the non-strict inequality $\beta' = (y \leq 1)$.
The skeleton phase of the {\sc conversion} procedure behaves
exactly as shown before, producing closure points $c_4$ and $c_5$.
We then process $\ns \in \NS^\pm$ in the {\sc move-ns} function:
\begin{align*}
  \ns^*
    &= \proj^{\beta'}_\SK\bigl( \suppcl_\SK(\ns) \bigr) \\
    &= \suppcl_\SK\bigl( \{ c_0, c_3 \} \bigr) \inters \SK^0 \\
    &= \{ c_0, c_3, c_4 \} \inters \{ c_4, c_5 \} \\
    &= \{ c_4 \}.
\end{align*}
Since $\ns^\star$ is a singleton, it will be upgraded to become
a skeleton point by procedure {\sc promote-singletons},
thereby obtaining the new skeleton component
$\SK = \langle \emptyset, \emptyset, C, \SP \rangle$,
where $C = \{ c_0, c_1, c_5 \}$ and $\SP = \{ c_4 \}$,
and the new non-skeleton component $\NS = \emptyset$.
Hence, we obtain the polyhedron in the right hand side of the figure;
note that the skeleton point $c_4$ is responsible for the inclusion
of the facets $(c_0, c_4]$ and $[c_4, c_5)$ in the polyhedron.
\end{example}

\subsubsection*{Creating new supports.}

On the one hand, the choice of representing only the minimal elements
of the upward closed set $\NS$ enables many efficiency improvements;
on the other hand, it also means that some care has to be taken
before removing these minimal elements.

As an example, consider the case of a support $\ns \in \NS^-$
when dealing with a non-strict inequality constraint $\beta$:
this support is going to be removed from $\NS$
in \cref{conv.remove-ns} of the {\sc conversion} function.
However, by doing so, we are also implicitly removing
other supports from the set $\upcl \ns$,
here included some supports that do not belong to $\NS^-$
and hence should be kept in $\NS$.
Therefore, at each iteration,
we have to explore the set of filled faces
and detect the ones that are going to lose their filler:
the corresponding minimal supports will be added to $\NS^\star$.
Moreover, when processing a non-strict inequality constraint,
we also need to consider the new faces introduced by the
constraint: the corresponding supports can be found
by projecting on the constraint hyperplane those faces that
are possibly filled by an element in $\SP^+$ or $\NS^+$.

\input{pseudo-create}

This is the task of the {\sc create-ns} function,
shown in Pseudocode~\ref{alg:create}.
This function uses {\sc enumerate-faces} as a helper:%
\footnote{This enumeration phase is inspired by
the algorithm in~\cite{KaibelP02}.}
the latter provides an enumeration of all the (higher dimensional)
faces that contain the initial support $\ns$.
The new faces are obtained
by adding to $\ns$ a new generator $g$
and then composing the projection and support closure
functions, as done in function {\sc move-ns}.

For efficiency purposes,
in function {\sc create-ns} a case analysis is performed
so as to suitably restrict the search area of the enumeration phase.
Since the faces we are going to compute have to be projected,
it is enough to consider those that can cross the constraint:
hence, when adding a new generator $g$ to a non-skeleton support $\ns$,
we consider only those coming from the opposite side of the constraint
(for instance, when processing $\ns \in \NS^-$ we consider $g \in \SK^+$,
disregarding the generators in $\SK^-$ and $\SK^0$).
We also avoid adding a point to $\ns$,
since this would definitely yield a redundant support.

\begin{figure}
\centering
{\scriptsize
\includegraphics{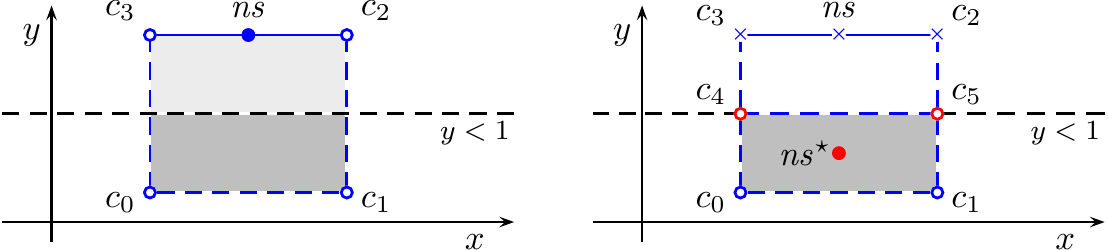}
}
\caption{Application of {\sc create-ns}
when adding a strict inequality.}
\label{fig:s-create}
\end{figure}

\begin{example}
Consider the polyhedron $\cP \in \Pset_2$ on the left hand side
of Figure~\ref{fig:s-create}.
The skeleton $\SK = \langle \emptyset, \emptyset, C, \emptyset \rangle$
is composed by the four closure points in $C = \{ c_0, c_1, c_2, c_3 \}$;
the non-skeleton $\NS = \{ \ns \}$ contains a single support
$\ns = \{ c_2, c_3 \}$, which makes sure that
the open segment $(c_2, c_3)$ is included in $\cP$.
By upward closure, this non-skeleton point is also
the filler for the whole polyhedron;
in particular, it fills $\relint(\cP)$.

The strict inequality makes $\ns \in \NS^-$, since all the
generators in the support are in $\SK^-$;
hence, support $\ns$ is processed by \cref{create.enum-ns-neg}
of function {\sc create-ns}.
The call to function {\sc enumerate-faces} will produce
new supports by adding to $\ns$ a generator from $\SK^+$
and then computing the corresponding support closure and projection.
Namely, it will compute
\begin{align*}
  \proj^\beta_\SK\bigl( \suppcl_\SK(\ns \union \{ c_0 \} ) \bigr)
    &= \suppcl_\SK\bigl(\ns \union \{ c_0 \} \bigr) \setdiff \SK^- \\
    &= \{ c_0, c_1, c_2, c_3, c_4, c_5 \} \setdiff \{ c_2, c_3 \} \\
    &= \{ c_0, c_1, c_4, c_5 \}, \\
  \proj^\beta_\SK\bigl( \suppcl_\SK(\ns \union \{ c_1 \} ) \bigr)
    &= \suppcl_\SK\bigl(\ns \union \{ c_1 \} \bigr) \setdiff \SK^- \\
    &= \{ c_0, c_1, c_2, c_3, c_4, c_5 \} \setdiff \{ c_2, c_3 \} \\
    &= \{ c_0, c_1, c_4, c_5 \}.
\end{align*}
Hence, the new (minimal) support $\ns^\star = \{ c_0, c_1, c_4, c_5 \}$
will be added to $\NS^\star$.
The resulting polyhedron, shown in the right hand side of the figure,
is described by the skeleton
\(
  \SK = \langle
          \emptyset, \emptyset,  \{ c_0, c_1, c_4, c_5 \}, \emptyset
        \rangle
\)
and the non-skeleton $\NS = \{ \ns^\star \}$.
\end{example}

\begin{figure}
\centering
{\scriptsize
\includegraphics{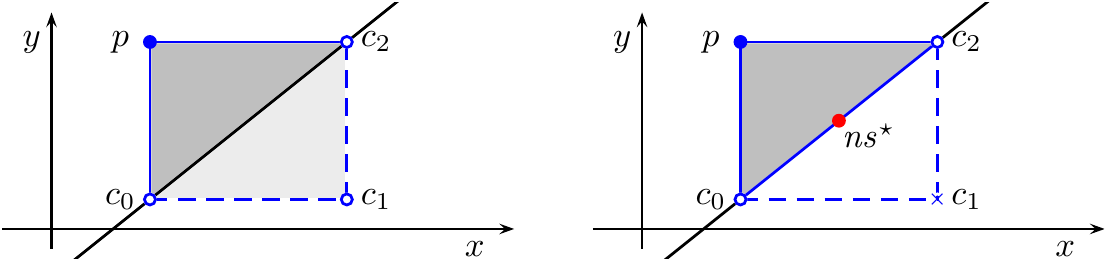}
}
\caption{Application of {\sc create-ns}
when adding a non-strict inequality.}
\label{fig:ns-create}
\end{figure}

\begin{example}
\label{ex:ns-create}
Consider polyhedron $\cP \in \Pset_2$ in the left hand side
of Figure~\ref{fig:ns-create}, described by skeleton
$\SK = \langle \emptyset, \emptyset, \{ c_0, c_1, c_2 \}, \{ p \} \rangle$
and non-skeleton $\NS = \emptyset$.
The partition for $\SK$ induced by the non-strict inequality is as follows:
\begin{align*}
  \SK^+ &= \langle \emptyset, \emptyset, \emptyset, \{ p \} \rangle, \\
  \SK^0 &= \langle \emptyset, \emptyset, \{ c_0, c_2 \}, \emptyset \rangle, \\
  \SK^- &= \langle \emptyset, \emptyset, \{ c_1 \}, \emptyset \rangle.
\end{align*}
There are no adjacent generators in $\SK^+$ and $\SK^-$,
so that the call to function
`$\combineadj{\beta}(\SK^+, \SK^-)$'
on \cref{conv.sk-comb-b} of {\sc conversion}
leaves $\SK^\star$ empty.

When processing the non-skeleton component,
the skeleton point in $\SK^+$ will be considered in
\cref{create.enum-sp-pos} of function {\sc create-ns}.
The corresponding call to function {\sc enumerate-faces}
produces new supports by first adding to $\{ p \}$ each generator in $\SK^-$
and then computing the corresponding support closure and projection.
Namely, it will compute
\begin{align*}
  \ns^\star
    & = \proj^\beta_\SK\bigl( \suppcl_\SK(\{ p \} \union \{ c_1 \} ) \bigr) \\
    &= \suppcl_\SK\bigl( \{ c_1, p \} \bigr) \inters \SK^0 \\
    &= \{ c_0, c_1, c_2, p \} \inters \{ c_0, c_2 \} \\
    &= \{ c_0, c_2 \},
\end{align*}
thereby producing the filler for the open segment $(c_0, c_2)$.
The resulting polyhedron, shown in the right hand side of the figure,
is thus described by the skeleton
$\SK = \langle \emptyset, \emptyset, \{ c_0, c_2 \}, \{ p \} \rangle$
and the non-skeleton $\NS = \{ \ns^\star \}$.
\end{example}

It is worth noting that, when handling Example~\ref{ex:ns-create}
adopting an entirely geometrical representation (as done in~\cite{Perri12th}),
closure point $c_1$ needs to be geometrically combined with point $p$
even if these two generators are \emph{not} adjacent.
In general, this leads to a significant efficiency penalty.
Similarly, an implementation based on the $\epsilon$-representation
will have to geometrically combine closure point $c_1$ with point $p$
(and/or with some other $\epsilon$-redundant points), because
the addition of the slack variable makes them adjacent.

In contrast, an implementation based on the new approach
is going to obtain a twofold benefit:
first, the distinction between the skeleton and non-skeleton
components allows for restricting the handling of non-adjacent
combinations to the non-skeleton phase,
thereby recovering the corresponding optimizations on the skeleton part;
second, by exploiting the combinatorial representation,
the non-skeleton component can be processed by using
set index operations only,
i.e., computing no linear combination at all.
As a consequence, the implementation is able to correctly
deal with closure points and strict inequalities without
a significant increase in the number of computationally heavy operations.

\subsubsection*{Handling special cases.}

In the previous paragraphs we have provided an explanation
of the core of the {\sc conversion} function.
We conclude by briefly discussing
those portions of Pseudcode~\ref{alg:conversion}
that are meant to efficiently handle some special cases.
Note that, being just optimizations,
these portions could be removed without compromising correctness.

\input{pseudo-misc}

In \crefrange{conv.line-b}{conv.line-e} of {\sc conversion}
we consider the case when constraint $\beta$ is violated
by a line. This special case is handled in procedure
{\sc violating-line} in Pseudocode~\ref{alg:line}.
The pseudocode is similar to the corresponding special case
for topologically closed polyhedra except that,
when processing a strict inequality constraint,
the helper procedure {\sc strict-on-eq-points} gets called:
this can be seen as a tailored version of the {\sc create-ns}
function, also including the final updating of $\SK$ and $\NS$.

In \crefrange{conv.spec-b}{conv.spec-e} of {\sc conversion}
we consider instead the cases
when $\SK^+$ or $\SK^-$ (or both) are empty.
Here we perform a few additional checks to see if an inconsistency
has been detected, making the polyhedron empty and thereby
allowing for an early exit from the main loop.
If this is not the case,
we efficiently update the $\SK$ and $\NS$ components,
possibly calling helper procedure {\sc strict-on-eq-points}.

%% file: pseudo-conv.tex
\begin{algorithm}
\caption{Incremental conversion from constraints to generators.}
\label{alg:conversion}
\begin{algorithmic}[2]
\Function{conversion}{$\cC_{\mathit{in}}$, $\langle \SK, \NS \rangle$}
\ForAll {$\beta \in \cC_{\mathit{in}}$}
  \State skel\_partition($\beta$, $\SK$);  \label{conv.sk-part}
  \State nonskel\_partition($\langle \SK, \NS \rangle$);
    \label{conv.ns-part}
  \If {line $\vect{l} \in \SK^+ \union \SK^-$}    \label{conv.line-b}
    \Comment{$\beta$ violates line $\vect{l}$}
    \State \Call{violating-line}{$\beta$, $\vect{l}$,
                                 $\langle \SK, \NS \rangle$};
    \label{conv.line-e}

  \ElsIf {$\SK^- = \emptyset$}    \label{conv.spec-b}
    \If {is\_equality($\beta$)}
      \If {$\SK^0 = \emptyset$}
        \State \Return $\langle \emptyset, \emptyset \rangle$;
               \Comment{$\cP$ is empty}
      \Else
        \State $\langle \SK, \NS \rangle \gets \langle \SK^0, \NS^0 \rangle$;
      \EndIf
    \ElsIf {is\_strict\_ineq($\beta$)}
      \If {$\SK^+ = \emptyset$}
        \State \Return $\langle \emptyset, \emptyset \rangle$;
               \Comment{$\cP$ is empty}
      \ElsIf {$\SK^0 \neq \emptyset$}
        \State \Call{strict-on-eq-points}{$\beta$, $\langle \SK, \NS \rangle$};
      \EndIf
    \EndIf

  \ElsIf {$\SK^+ = \emptyset$}
    \If {is\_strict\_ineq($\beta$) \textbf{or} $\SK^0 = \emptyset$}
      \State \Return $\langle \emptyset, \emptyset \rangle$;
             \Comment{$\cP$ is empty}
    \Else
      \State $\langle \SK, \NS \rangle \gets \langle \SK^0, \NS^0 \rangle$;
    \EndIf
    \label{conv.spec-e}

  \Else \Comment{$\SK^+ \neq \emptyset$ \textbf{and} $\SK^- \neq \emptyset$}
    \State $\SK^\star \gets \combineadj{\beta}(\SK^+, \SK^-)$;
      \label{conv.sk-comb-b}
    \State $\SK^0 \gets \SK^0 \union \SK^\star$; \label{conv.sk-comb-e}
    \State $\NS^\star \gets$
           \Call{move-ns}{$\beta$, $\langle \SK, \NS \rangle$};
           \label{conv.move-ns}
    \State $\NS^\star \gets \NS^\star \union$
           \Call{create-ns}{$\beta$, $\langle \SK, \NS \rangle$};
           \label{conv.create-ns}

    \If {is\_equality($\beta$)} \label{conv.final-b}
      \State
        \(
          \langle \SK, \NS \rangle
            \gets
          \langle \SK^0, \NS^0 \oplus \NS^\star \rangle
        \);
    \ElsIf {is\_nonstrict\_ineq($\beta$)}
      \State
        \(
          \langle \SK, \NS \rangle
            \gets
          \langle
            \SK^+ \union \SK^0,
            (\NS^+ \union \NS^0) \oplus \NS^\star
          \rangle
        \); \label{conv.remove-ns}
    \Else \Comment{is\_strict\_ineq($\beta$)}
      \State $\SK^0 \gets \pbecomecp(\SK^0)$;
      \label{conv.p-to-cp}
      \State
        \(
          \langle \SK, \NS \rangle
            \gets
          \langle
            \SK^+ \union \SK^0,
            \NS^+ \oplus \NS^\star
          \rangle
        \);
    \EndIf \label{conv.final-e}
    \State \Call {promote-singletons}{$\langle \SK, \NS \rangle$};
    \label{conv.promote-sing}
  \EndIf
\EndFor \Comment{end of loop on $\cC_{\mathit{in}}$}
\State \Return{$\langle \SK, \NS \rangle$};
\EndFunction
\end{algorithmic}
\end{algorithm}

%% file: pseudo-promote.tex
\begin{algorithm}
\caption{Helper procedure for promoting singleton supports.}
\label{alg:promote}
\begin{algorithmic}
\Procedure{promote-singletons}{$\langle \SK, \NS \rangle$}
\State let $\SK = \langle L, R, C, \SP \rangle$;
\ForAll
  {$\ns \in \NS$ such that
   $\ns = \langle \emptyset, \emptyset, \{ \vect{c} \}, \emptyset \rangle$}
  \State $\NS \gets \NS \setdiff \{ \ns \}$;
  \State $C \gets C \setdiff \{ \vect{c} \}$;
  \State $\SP \gets \SP \union \{ \vect{c} \}$;
\EndFor
\EndProcedure
\end{algorithmic}
\end{algorithm}

%% file: pseudo-move.tex
\begin{algorithm}
\caption{Helper function for moving supports.}
\label{alg:move}
\begin{algorithmic}[2]
\Function{move-ns}{$\beta$, $\langle \SK, \NS \rangle$}
\State $\NS^\star \gets \emptyset$;
\ForAll {$\ns \in \NS^\pm$}
  \State
    \(
      \NS^\star \gets \NS^\star \union \{ \proj^\beta_\SK(\suppcl_\SK(\ns)) \}
    \); \label{move.proj-cl}
\EndFor
\State \Return $\NS^\star$;
\EndFunction
\end{algorithmic}
\end{algorithm}

%% file: pseudo-create.tex
\begin{algorithm}
\caption{Helper functions for creating new supports.}
\label{alg:create}
\begin{algorithmic}[2]
\Function{create-ns}{$\beta$, $\langle \SK, \NS \rangle$}
\State $\NS^\star \gets \emptyset$;
\State let $\SK = \langle L, R, C, \SP \rangle$;
\ForAll {$\vect{p}$ in $\SP^-$}
  \State $\NS^\star \gets \NS^\star \union$
         \Call{enumerate-faces}{$\beta$, $\{\vect{p}\}$, $\SK^+$, $\SK$};
         \label{create.enum-sp-neg}
\EndFor
\ForAll {$\ns \in \NS^-$}
  \State $\NS^\star \gets \NS^\star \union$
         \Call{enumerate-faces}{$\beta$, $\ns$, $\SK^+$, $\SK$};
         \label{create.enum-ns-neg}
\EndFor

\If {is\_strict\_ineq($\beta$)}
  \ForAll {$\vect{p} \in \SP^0$}
    \State $\NS^\star \gets \NS^\star \union$
           \Call{enumerate-faces}{$\beta$, $\{ \vect{p} \}$, $\SK^+$, $\SK$};
           \label{create.enum-sp-zero}
  \EndFor
  \ForAll {$\ns \in \NS^0$}
    \State $\NS^\star \gets \NS^\star \union$
           \Call{enumerate-faces}{$\beta$, $\ns$, $\SK^+$, $\SK$};
           \label{create.enum-ns-zero}
  \EndFor

\ElsIf {is\_nonstrict\_ineq($\beta$)}
  \ForAll {$\vect{p} \in \SP^+$}
    \State $\NS^\star \gets \NS^\star \union$
           \Call{enumerate-faces}{$\beta$, $\{ \vect{p} \}$, $\SK^-$, $\SK$};
           \label{create.enum-sp-pos}
  \EndFor
  \ForAll {$\ns \in \NS^+$}
    \State $\NS^\star \gets \NS^\star \union$
           \Call{enumerate-faces}{$\beta$, $\ns$, $\SK^-$, $\SK$};
           \label{create.enum-ns-pos}
  \EndFor
\EndIf
\State \Return $\NS^\star$;
\EndFunction
\end{algorithmic}

\begin{algorithmic}[2]
\Function{enumerate-faces}{$\beta$, $\ns$, $\SK'$, $\SK$}
\State $\NS^\star \gets \emptyset$;
\State let $\SK' = \langle L', R', C', \SP' \rangle$;
\ForAll {$g \in (R' \union C')$}
  \State $\ns' \gets \ns \union \{ g \}$;
  \State
    $\NS^\star \gets \NS^\star \union \{ \proj^\beta_\SK(\suppcl_\SK(\ns')) \}$;
\EndFor
\State \Return $\NS^\star$;
\EndFunction
\end{algorithmic}

\end{algorithm}

%% file: pseudo-misc.tex
\begin{algorithm}
\caption{Processing a line violating constraint $\beta$.}
\label{alg:line}
\begin{algorithmic}[2]
\Procedure{violating-line}{$\beta$, $\vect{l}$, $\langle \SK, \NS \rangle$}
\State split $\vect{l}$ into rays $\vect{r}^+$ satisfying $\beta$
       and $\vect{r}^-$ violating $\beta$;
\State $\vect{l} \gets \vect{r}^+$;
\ForAll {$g \in \SK$}
  \State $g \gets \combine{\beta}(g, \vect{l})$;
\EndFor
\Comment{now $\vect{l} \in \SK^+$ and all other $g \in \SK^0$}
\If {is\_equality($\beta$)}
  \State $\SK \gets \SK^0$;
\ElsIf {is\_strict\_ineq($\beta$)}
  \State \Call{strict-on-eq-points}{$\beta$, $\langle \SK, \NS \rangle$};
\EndIf
\EndProcedure
\end{algorithmic}
\end{algorithm}

\begin{algorithm}
\caption{Processing points saturating a strict inequality.}
\label{alg:process-eq}
\begin{algorithmic}[2]
\Procedure{strict-on-eq-points}{$\beta$, $\langle \SK, \NS \rangle$}
\State $\NS^\star \gets \emptyset$;
\State let $\SK^0 = \langle L^0, R^0, C^0, \SP^0 \rangle$;
\ForAll {$\vect{p} \in \SP^0$}
  \State $\NS^\star \gets \NS^\star \union$
         \Call{enumerate-faces}{$\beta$, $\{ \vect{p} \}$, $\SK^+$, $\SK$};
\EndFor
\ForAll {$\ns \in \NS^0$}
  \State $\NS^\star \gets \NS^\star \union$
         \Call{enumerate-faces}{$\beta$, $\ns$, $\SK^+$, $\SK$};
\EndFor
\State $\SK^0 \gets$ points-become-closure-points($\SK^0$);
\State
  \(
    \langle \SK, \NS \rangle
      \gets
        \langle \SK^+ \union \SK^0, \NS^+ \oplus \NS^\star \rangle
  \);
\EndProcedure
\end{algorithmic}
\end{algorithm}

%% file: duality.tex
The definitions and observations given in Section~\ref{sec:new-repr}
for a geometric generator system have their dual versions
working on a geometric \emph{constraint} system.
In the following we provide a brief overview of these correspondences,
which are also summarized in Table~\ref{tab:duality}.

\begin{table}
\centering
\begin{tabular}{ccc}
\cmidrule{2-3}
\multicolumn{1}{c}{} &
Generators & Constraints \\
\toprule
\textbf{Geometric skeleton}&&\\
singular & line & equality \\
non-singular & ray or closure point & non-strict inequality \\
semantics & $\gen(\SK) = \emptyset$ & $\con(\SK) = \cl(\cP)$ \\
\midrule
\textbf{Combinatorial non-skeleton}&&\\
abstracts & point & strict inequality \\
element role & face filler & face cutter \\
represents & upward closed set & downward closed set \\
encoding & minimal support & minimal support \\
singleton & skeleton point & skeleton strict inequality \\
\bottomrule
\multicolumn{3}{c}{}\\
\end{tabular}
\caption{Correspondences between generator and constraint concepts.}
\label{tab:duality}
\end{table}

For a non-empty $\cP = \con(\cC) \in \Pset_n$,
the skeleton component of the geometric constraint system
$\cC = \langle C_=, C_\geq, C_> \rangle$
includes the non-redundant constraints defining
the topological closure $\cQ = \cl(\cP)$.
Denoting by $\SC_>$ the set of \emph{skeleton strict inequalities}
(i.e., those in $C_>$ whose corresponding non-strict inequality
is not redundant for $\cQ$), we can define
\(
  \SK_\cQ \defeq \langle C_=, C_\geq \union \SC_> , \emptyset \rangle
\),
so that $\cQ = \con(\SK_\cQ)$.

The \emph{ghost} faces of $\cP$ are the faces of the topological
closure $\cQ$ that do not intersect $\cP$:
\[
  \GFaces_\cP
    \defeq
      \{\, F \in \CFaces_\cQ \mid F \inters \cP = \emptyset \,\};
\]
as a consequence, we obtain
\[
  \cP = \con(\SK_\cQ) \setdiff \bigunion \GFaces_\cP.
\]
With the only exception of the empty face,
the elements in $\GFaces$ are exactly those
not occurring in $\cl(\NNCFaces)$.
The set $\GFaces' \defeq \GFaces \union \{ \cQ \}$
is a meet sublattice of $\CFaces$;
moreover, $\GFaces$ is downward closed and thus can be efficiently
represented by its \emph{maximal} elements
(with respect to the set inclusion relation on faces),
which are the dual-atoms of $\GFaces'$.

The skeleton support $\SK_F$ of a face $F \in \CFaces_\cQ$ is defined
as the set of all the skeleton constraints that are saturated
by all the points in $F$.
Each face $F \in \GFaces$ saturates a strict inequality
$\beta_> \in C_>$: we can represent such a face using its skeleton
support $\SK_F$ of which $\beta_>$ is a possible materialization.
A constraint system non-skeleton component $\NS \sseq \NSset$
is thus a combinatorial representation of the \emph{strict inequalities}
of the polyhedron.

Hence, the non-skeleton components for generators and constraints
have a complementary role:
in the case of generators they are face \emph{fillers},
marking the minimal faces that are \emph{included} in $\NNCFaces$;
in the case of constraints they are face \emph{cutters},
marking the maximal faces that are \emph{excluded} from $\NNCFaces$.
Note however that, when representing a cutter in $\GFaces$
using its skeleton support
the non-redundant cutters
are again those having a \emph{minimal} skeleton support,
as is the case for the fillers.

As it happens with lines,
all the equalities in $\cC_=$ are included in all the supports
$\ns \in \NS$ so that, for efficiency, they are not represented explicitly.
After removing the equalities,
a singleton $\ns = \{ \beta \} \in \NS$ stands for
a \emph{skeleton strict inequality} constraint,
which is better represented in the skeleton component,
thereby obtaining $\SK = \langle C_=, C_\geq, \SC_> \rangle$.
Hence, a support $\ns \in \NS$ is redundant if there exists
$\ns' \in \NS$ such that $\ns' \subset \ns$ or if
$\ns \inters \SC_> \neq \emptyset$.

The handling of the empty face deserves a technical observation
(which can be skipped when adopting a higher level point of view).
The empty face is always cut away from the polyhedron,
hence it belongs to $\GFaces$ even when $\cP$ is topologically closed.
The skeleton support for the empty face can be given by
a set of skeleton constraints whose hyperplanes have an empty intersection
or by a constraint that is saturated by no points or closure points:
the latter happens to be the case for the positivity constraint `$1 \geq 0$'
(see Section~\ref{sec:prelims}).
It follows that, when the positivity constraint is not redundant,
the empty face should be represented by the non-skeleton support
$\ns = \{ 1 \geq 0 \}$; being a singleton, this will be promoted
into the skeleton component, thereby encoding the positivity constraint
as a \emph{strict} inequality `$1 > 0$'.
Otherwise, when the positivity constraint is redundant,
the empty face will be cut by the maximal support $\ns = C_\geq$.

When the concepts underlying the skeleton and non-skeleton
representation are reinterpreted as discussed above,
it is possible to define a conversion procedure mapping
a generator representation into a constraint representation
which is very similar to the one from constraints to generators
shown in Section~\ref{sec:conv}.
One of the few differences, only occurring when performing
a non-incremental conversion, can be seen in the \emph{initialization} phase.
While in Section~\ref{sec:conv} we are starting
from a representation of the universe polyhedron,
having preprocessed the positivity constraint only,
when converting from generators to constraints
we look for a point in $\cG_{\mathit{in}}$: if such a point does not exists,
the polyhedron is empty; otherwise, we preprocess it to obtain
a skeleton constraint system being made of $n$ linear equality constraints
(plus the strict positivity constraint).
Another difference is in the handling of the special cases
in \crefrange{conv.spec-b}{conv.spec-e} of {\sc conversion}.
When converting from generators to constraints, since we incrementally
add new generators to a non-empty polyhedron, there is no way
we can obtain an inconsistency: hence, the checks corresponding
to the comments `$\cP$ is empty' can be omitted.
The rest of the code is almost unchanged:
as a matter of fact,
for each of the functions and procedures
in Pseudocodes~\ref{alg:promote},~\ref{alg:move},%
~\ref{alg:create},~\ref{alg:line} and~\ref{alg:process-eq},
the corresponding {\Cplusplus} implementation is based on
either a single function or a function template
(which is then instantiated for both cases).

%% file: exp-eval.tex
The new representation and conversion algorithms for NNC polyhedra
presented in the previous sections have been implemented and
tested, for both correctness and efficiency,
in the context of the Parma Polyhedra Library (PPL).%
\footnote{All experiments have been performed on a laptop
with an Intel Core i7-3632QM CPU, 16 GB of RAM and
running GNU/Linux 4.13.0-16.}

Due to the adoption of the direct encoding for constraints and generators,
a full integration of the new algorithm in the domain of NNC polyhedra
provided by the PPL is not possible, since the latter assumes
the presence of the slack variable $\epsilon$.
Rather, the approach adopted is to intercept every call
to the PPL's conversion procedures
(working on the $\epsilon$-representations in $\CPset_{n+1}$)
and pair it with a corresponding call to the newly defined
conversion algorithms (working on the new representations in $\Pset_n$).

\tikzset{
  block/.style    = { rectangle, draw, text centered },
  iodata/.style   = { block, text width=12em },
  algstep/.style  = { block, rounded corners },
  galgstep/.style = { algstep, text width=7em, fill=green!30 },
  balgstep/.style = { algstep, text width=8em, fill=blue!30 }
}

\begin{figure}
\begin{centering}
\begin{tikzpicture}
\matrix(m)[matrix of nodes,
           column sep=0.5cm, row sep=4mm,
           align=center,
           nodes={rectangle, draw, anchor=center}]
{
  |[iodata]| {$\epsilon$-repr $\cC_{\mathit{in}}$ (resp., $\cG_{\mathit{in}}$)}
    & & \\
    & |[galgstep]| {$\epsilon$-less encoding} & \\
    & & |[iodata]| {$\epsilon$-less $\cC'_{\mathit{in}}$
                    (resp., $\cG'_{\mathit{in}}$)} \\
  |[balgstep]| {init DD}    & & |[balgstep]| {new init DD} \\
  |[balgstep]| {conversion} & & |[balgstep]| {new conversion} \\
  |[balgstep]| {simplify}   & & |[balgstep]| {new simplify} \\
  |[iodata]| {$\epsilon$-repr DD \\
              $( \cC_{\mathit{out}}, \cG_{\mathit{out}} )$}
    & & |[iodata]| {skel/non-skel DD \\
                    \(
                      \bigl(
                        \langle \cC_\SK, \cC_\NS \rangle,
                        \langle\cG_\SK, \cG_\NS \rangle
                      \bigr)
                    \)} \\
    & |[galgstep]| {equiv. and \\ non-redund. \\ checks} & \\
};
\path [>=latex,->] (m-1-1.south) edge [out=-90, in=-180] (m-2-2.west);
\path [>=latex,->] (m-2-2.east) edge [out=0 in=60] (m-3-3.north);
\path [>=latex,->] (m-1-1) edge (m-4-1);
\path [>=latex,->] (m-4-1) edge (m-5-1);
\path [>=latex,->] (m-5-1) edge (m-6-1);
\path [>=latex,->] (m-6-1) edge (m-7-1);
\path [>=latex,->] (m-3-3) edge (m-4-3);
\path [>=latex,->] (m-4-3) edge (m-5-3);
\path [>=latex,->] (m-5-3) edge (m-6-3);
\path [>=latex,->] (m-6-3) edge (m-7-3);
\path [>=latex,->] (m-7-1.south) edge [out=-90, in=-180] (m-8-2.west);
\path [>=latex,->] (m-7-3.south) edge [out=-90, in=0] (m-8-2.east);
\end{tikzpicture}
\end{centering}
\caption{High level diagram for the experimental evaluation
(non-incremental case).}
\label{fig:test-strategy}
\end{figure}
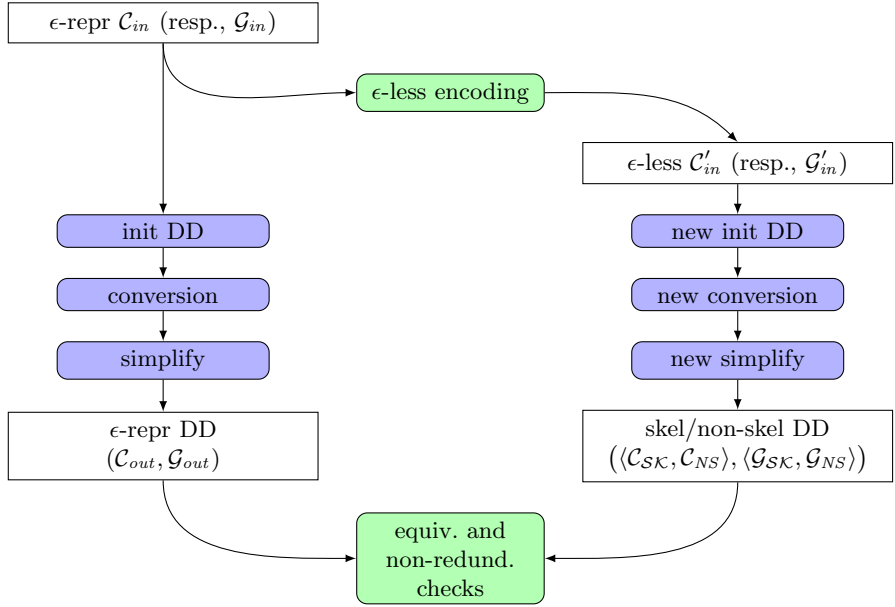


The diagram in Figure~\ref{fig:test-strategy},
where we consider the case of a \emph{non-incremental} conversion,
provides a more detailed description of the experimental setting.
On the left hand side of the diagram we see the application
of the standard PPL conversion procedure:
the input constraint system $\cC_{\mathit{in}}$
(resp., generator system $\cG_{\mathit{in}}$)
for the $\epsilon$-representation of the NNC polyhedron
is processed by the three computational phases
(`init DD', `conversion' and `simplify')
so as to produce the output $\epsilon$-representation DD pair
$(\cC_{\mathit{out}}, \cG_{\mathit{out}})$.
A copy of the input system is processed by the `$\epsilon$-less encoding'
phase so as to remove the slack variable and produce a corresponding
$\epsilon$-less version $\cC'_{\mathit{in}}$ (resp., $\cG'_{\mathit{in}}$);
this is processed by the three computational phases of the new algorithm
(`new init DD', `new conversion' and `new simplify')
to produce the output DD pair, which is based
on the new skeleton/non-skeleton representation
$\bigl( \langle \cC_\SK, \cC_\NS \rangle, \langle \cG_\SK, \cG_\NS \rangle \bigr)$.
After both the old and new conversions are completed,
the two outputs are passed to a checking phase,
where the new output is tested for both semantic equivalence
and non-redundancy.

A similar diagram could be shown for an \emph{incremental} conversion:
in this case, the input is a DD pair for an $\epsilon$-representation
together with some new constraints/generators to be processed by
the standard conversion phases (skipping the `init DD' phase).
The `$\epsilon$-less encoding' phase translates all the inputs
into the corresponding $\epsilon$-less representations,
including an input skeleton/non-skeleton DD pair,
to be processed by the new algorithms
(again, skipping the `new init DD' phase).

As far as correctness is concerned,
the final checking phase was successful on all the experiments performed,
which includes all of the tests present in the PPL library itself,
as well as several new tests explicitly written to stress specific portions
of the new algorithms.

In order to assess the efficiency of the new algorithm,
additional code was added so as to measure the time spent
inside the standard and new computational phases,
disregarding the input encoding and output checking phases.

The first experiment on efficiency is meant to evaluate the \emph{overhead}
incurred by the new representation and algorithm for NNC polyhedra
when processing topologically closed polyhedra,
so as to compare it with the corresponding overhead incurred
by the $\epsilon$-representation.
To this end, we considered the \texttt{ppl\_lcdd} demo application
of the Parma Polyhedra Library,
which solves the \emph{vertex/facet enumeration problem}.
In Table~\ref{tab:c-vs-nnc} we report the results obtained on
a selection%
\footnote{We only show those tests where the absolute difference
between the PPL closed polyhedron time and the new algorithm time is
bigger than 10 milliseconds.}
of the test benchmarks,
whose name is reported in the first column of the table.
Note that, for each benchmark, the application performs a single
conversion of representation, taking as input a system of constraints
(for those tests having `.ine' as file extension)
or a system of generators (for those tests having `.ext' as extension).
For each of these tests we show the efficiency measures obtained
in the following cases:
when using the standard conversion algorithm for closed polyhedra
(columns 2--4);
when using the standard conversion algorithm for the $\epsilon$-representation
of NNC polyhedra (columns 5--7);
and when using the new conversion algorithm for the new representation
of NNC polyhedra (columns 8--10).
The three values measured are:
\begin{description}
\item[time]
the time spent in the considered computation phase, in milliseconds;
\item[vec ops]
the number of vector operations computed
(scalar products and linear combinations), in thousands;
\item[sat ops]
the number of saturation row operations computed
(bit-vector population counts, unions and inclusion tests),
in millions.
\end{description}
Also note that in each case we report, in different rows, two sets of values:
the first row shows the results for the `conversion' phase,
while the second row shows the results for the `simplify' phase;
the latter is shown just to stress that it is usually negligible,
since most of the computation time is spent in the `conversion' phase proper.

\begin{table}
\begin{center}
{\mytablesize
\begin{tabular}{||l||r|r|r||r|r|r||r|r|r||}
\hhline{|t:=:t:===:t:===:t:===:t|}
\multicolumn{1}{||c||}{test}
  & \multicolumn{3}{c||}{closed poly}
  & \multicolumn{3}{c||}{$\epsilon$-repr NNC}
  & \multicolumn{3}{c||}{$\langle \SK, \NS \rangle$ NNC} \\
\hhline{~||-|-|-||-|-|-||-|-|-||}
  & time & vec ops & sat ops
  & time & vec ops & sat ops
  & time & vec ops & sat ops \\
\hhline{|:=::===::===::===:|}
cp6.ext
  &   24 &   6.4 &  1.1 &   52 &   14.1 &   5.3 &  12 &   6.4 &  1.1 \\
  &    0 &   --- &  0.0 &    0 &    --- &   0.0 &   0 &   --- &  0.0 \\
\hhline{||-||-|-|-||-|-|-||-|-|-||}
cross12.ine
  &   48 & 112.8 &  0.3 &  124 &  172.2 &   1.3 &  56 & 112.7 &  0.5 \\
  &  104 &   --- & 16.8 &  108 &    --- & 167.9 & 132 &   --- & 16.8 \\
\hhline{||-||-|-|-||-|-|-||-|-|-||}
in7.ine
  &   56 &   8.7 &  1.7 &  136 &   13.9 &   4.7 &  24 &   8.7 &  0.9 \\
  &    0 &   --- &  0.0 &    0 &    --- &   0.0 &   0 &   --- &  0.0 \\
\hhline{||-||-|-|-||-|-|-||-|-|-||}
kkd38\_6.ine
  &  656 &  64.7 & 28.3 & 2700 &  129.2 & 113.2 & 200 &  64.6 & 14.2 \\
  &    0 &   --- &  0.0 &    0 &    --- &   0.0 &   0 &   --- &  0.0 \\
\hhline{||-||-|-|-||-|-|-||-|-|-||}
kq20\_11\_m.ine
  &   56 &   8.7 &  1.7 &  132 &   13.9 &   4.7 &  24 &   8.7 &  0.9 \\
  &    0 &   --- &  0.0 &    0 &    --- &   0.0 &   0 &   --- &  0.0 \\
\hhline{||-||-|-|-||-|-|-||-|-|-||}
metric80\_16.ine
  &   44 &  20.9 &  2.3 &   84 &   32.1 &   5.4 &  24 &  20.4 &  2.0 \\
  &    0 &   --- &  0.0 &    0 &    --- &   0.0 &   0 &   --- &  0.0 \\
\hhline{||-||-|-|-||-|-|-||-|-|-||}
mit31-20.ine
  & 1308 &  69.4 & 88.7 & 5100 &  102.1 & 358.0 & 724 &  69.3 & 60.2 \\
  &    4 &   --- &  0.0 &    4 &    --- &   0.0 &  12 &   --- &  0.0 \\
\hhline{||-||-|-|-||-|-|-||-|-|-||}
mp6.ine
  &  100 &  35.1 &  6.4 &  260 &   60.3 &  17.6 &  68 &  38.4 &  8.0 \\
  &    0 &   --- &  0.0 &    0 &    --- &   0.0 &   0 &   --- &  0.0 \\
\hhline{||-||-|-|-||-|-|-||-|-|-||}
reg600-5\_m.ext
  &  956 & 725.3 & 24.3 & 3508 & 1460.9 & 117.7 & 688 & 725.3 & 12.9 \\
  &   16 &   --- &  0.4 &   40 &    --- &   1.4 &  44 &   --- &  0.4 \\
\hhline{||-||-|-|-||-|-|-||-|-|-||}
sampleh8.ine
  & 7184 & 543.8 & 307.4 & 28940 & 1086.7 & 1228.7 & 2904 & 543.8 & 153.8 \\
  &    8 &   --- &   0.0 &    16 &    --- &    0.0 &   32 &   --- &   0.0 \\
\hhline{||-||-|-|-||-|-|-||-|-|-||}
trunc10.ine
  & 1660 & 213.3 & 91.7 & 6928 &  423.8 & 396.6 & 784 & 212.8 & 89.9 \\
  &    0 &   --- &  0.0 &    0 &    --- &   0.0 &   0 &   --- &  0.0 \\
\hhline{|b:=:b:===:b:===:b:===:b|}
\end{tabular}
} 
\end{center}
\caption{Measuring the overhead of the conversion procedure
for NNC polyhedra; the topologically closed polyhedra
used as tests are part of the \texttt{ppl\_lcdd} test suite.
Units: time (ms), vec ops (K), sat ops (M).}
\label{tab:c-vs-nnc}
\end{table}

The inspection of the results in Table~\ref{tab:c-vs-nnc}
leads to a few observations.
As mentioned in Section~\ref{sec:old-repr},
the use of the $\epsilon$-representation for topologically closed
polyhedra incurs a significant overhead, which on the considered tests
ranges from 53\% (cross12.ine) to 317\% (trunc10.ine).
In contrast, the new representation and algorithm go beyond
all expectations: in almost all of the tests there is no overhead
at all (that is, any overhead incurred is so small to be masked by
the improvements obtained in other parts of the algorithm);
the efficiency gain ranges from 25\% (reg600-5\_m.ext)
to 70\% (kdd38\_6.ine);
the only slowdown, measuring 25\%,
is obtained on a test (cross12.ine)
where the time spent in the `simplify' phase
dominates the `conversion' phase.
It is worth stressing that we are comparing the time obtained
for the new algorithm for NNC polyhedra against the time obtained
by the standard algorithm for \emph{closed} polyhedra.
A direct comparison against the $\epsilon$-representation NNC polyhedra
results in much bigger efficiency gains (and no slowdown at all).

\begin{table}
\begin{center}
{\mytablesize
\begin{tabular}{||c||r||r|r|r||r|r||r|r||r||}
\hhline{|t:=:t:=:t:===:t:==:t:==:t:=:t|}
\multirow{2}{*}{algorithm}
  & \multicolumn{1}{c||}{iter}
  & \multicolumn{3}{c||}{iter repr sizes}
  & \multicolumn{2}{c||}{full conv}
  & \multicolumn{2}{c||}{incr conv}
  & \multicolumn{1}{c||}{time} \\
\hhline{||~||~||-|-|-||-|-||-|-||~||}
  & \multicolumn{1}{c||}{count}
  & avg & median & max
  & num & time & num & time
  & \multicolumn{1}{c||}{ratio} \\
\hhline{|:=::=::===::==::==::=:|}
$\epsilon$-repr standard
  & 1142 & 3381 & 3706 & 7259
  & 4 &    11 & 3 & 29800 & 1652.8 \\
$\epsilon$-repr enhanced
  &  525 &  169 &  109 & 1661
  & 7 &   240 & 0 &   --- &   13.3 \\
$\langle \SK, \NS \rangle$ standard
  &  314 &   56 &   62 &  156
  & 4 &     7 & 3 &    11 &    1.0 \\
\hhline{|b:=:b:=:b:===:b:==:b:==:b:=:b|}
\end{tabular}
} 
\end{center}
\caption{Comparison between the $\epsilon$-representation based
(standard and enhanced) computations for NNC polyhedra and
the one based on the new representations and conversion procedures.}
\label{tab:dual-hyp}
\end{table}

The second experiment is meant to evaluate the efficiency
gains obtained by the application of the new representation and algorithm
in a more appropriate context, i.e.,
when processing NNC polyhedra that are \emph{not} topologically closed.
To this end, we reconsider the same benchmark that was discussed
at length in~\cite[Table~2]{BagnaraHZ05FAC}:%
\footnote{The test \texttt{dualhypercubes.cc} is distributed with the
source code of the PPL.}
in this test, four NNC dual-hypercubes are combined by
a few convex polyhedral hull and intersection operations.
This test was meant to highlight the efficiency improvement
resulting from the adoption of an \emph{enhanced} evaluation strategy
(where a knowledgeable user of the library explicitly invokes,
when appropriate, the strong minimization procedures
for $\epsilon$-representations)
with respect to the \emph{standard} evaluation strategy
(where the user simply performs the required computation,
leaving the burden of optimization to the library developers).
In Table~\ref{tab:dual-hyp} we report the results obtained
for the most expensive test among those described
in~\cite[Table~2]{BagnaraHZ05FAC},
comparing the standard and enhanced evaluation strategies
for the $\epsilon$-representation (rows 1 and 2)
with the new algorithm (row 3).
For each algorithm, whose name is reported in column 1,
we show in column 2 the total number of iterations of
the conversion procedures and, in the next three columns,
the average, median and maximum sizes of the representations computed
at each iteration (i.e., the size of the intermediate results);
in columns from 6 to 9 we show the numbers of incremental
and non-incremental calls to the conversion procedures,
together with the corresponding time spent (in milliseconds);
in the final column, we show the overall time ratio,
computed with respect to the time spent by the new algorithm.

Even though adopting the standard computation strategy
(requiring no clever guess by the end user),
the new algorithm is able to outperform not only the standard,
but also the enhanced computation strategy for the $\epsilon$-representation.
As discussed in Section~\ref{sec:old-repr}, the reasons for
this efficiency improvement is that the enhanced computation
strategy is interfering with incrementality:
the figures in Table~\ref{tab:dual-hyp} confirm that
the new algorithm performs three of the seven
required conversions in an incremental way,
while in the enhanced case they are all non-incremental.
Moreover, a comparison of the iteration count and the size
of the intermediate results provides further evidence that
the new algorithm is able to maintain a non-redundant description
even \emph{during} the iterations of a conversion, which justifies
the impressive time improvements.

After having discussed the outcome of the experimental evaluation,
it is possible to highlight how the adoption of the new representation
and conversion procedure provides a solution for all of the issues
affecting the $\epsilon$-representation approach,
which were listed at the end of Section~\ref{sec:old-repr}.
\begin{enumerate}
\item
At the implementation level, no tricks are needed to hide the $\epsilon$
dimension, as in the new representation there is no slack variable at all.
\item
The overhead of the $\epsilon$-representation for generators
has simply disappeared: the skeleton points need not be matched
by corresponding closure points. This claim is backed up
by the efficiency results shown in Table~\ref{tab:c-vs-nnc}.
\item
The new conversion procedure is fully incremental:
it is able to remove the redundant elements from the representation
at each iteration of the main loop, thereby keeping the
intermediate results smaller. This claim is supported
by the efficiency results shown in Table~\ref{tab:dual-hyp}.
\end{enumerate}

%% file: concl.tex
We have presented a new approach for the representation
of NNC polyhedra in the Double Description framework.
The main difference of the new approach with respect to previous proposals
is that it adopts a direct representation, where the strict
inequality constraints and the closure points of NNC polyhedra
are encoded using no slack variable at all.
The new representation also distinguishes between the skeleton component,
which is encoded geometrically, and the non-skeleton component,
which is provided with a combinatorial encoding.

Based on this new representation, we have proposed and implemented
a variant of the Chernikova-like conversion procedure
which is able to achieve significant efficiency improvements
with respect to state-of-the-art implementations
of the domain of NNC polyhedra.

As future work, we plan to provide a full implementation of the
domain of NNC polyhedra which is based on this new representation
and conversion algorithm.
To this end, we will have to reconsider each semantic operator
already implemented by the existing libraries
(which are based on the addition of a slack variable),
so as to propose, implement and experimentally evaluate
a corresponding correct specification based on the new approach.

%% file: appendix.tex
\allowdisplaybreaks

We provide here proof sketches for the results stated
in Section~\ref{sec:new-repr}.


\begin{delayedproof}[Proposition~\ref{prop:skeleton}]
Let $\cG = \langle L, R, C, P \rangle$ and
consider a generator system
$\cG_m = \langle L_m, R_m, C_m, P_m \rangle$
in minimal form such that
$\gen(\cG_m) = \gen(\cG) = \cP$.
By Definition~\ref{def:skeleton}, we obtain
$\SK_\cQ = \langle L_m, R_m, C_m \union \SP_m, \emptyset \rangle$,
where $\SP_m \sseq P_m$ is the set of skeleton points in $P_m$.
Since each point $\vect{p} \in P_m \setdiff \SP_m$ can be obtained
by a combination of the generators in $L_m$, $R_m$ and $C_m \union \SP_m$,
we have the following chain of equivalences:
\begin{align*}
  \fullgen(\cG)
    &= \fullgen(\cG_m) \\
    &= \gen\bigl(
             \langle L_m, R_m, \emptyset , C_m \union P_m \rangle
           \bigr) \\
    &= \gen\bigl(
             \langle L_m, R_m, \emptyset , C_m \union \SP_m \rangle
           \bigr) \\
    &= \fullgen\bigl(
                 \langle L_m, R_m, C_m \union \SP_m, \emptyset \rangle
               \bigr) \\
    &= \fullgen(\SK_\cQ).
\end{align*}
Since function `$\fullgen$' interprets closure points as points,
it computes a topologically closed polyhedron, so that
$\fullgen(\SK_\cQ) = \cQ = \cl(\cP)$.
Moreover,
since $\SK_\cQ$ has been built from the generator system $\cG_m$
in minimal form, by construction it only keeps in $C_m \union \SP_m$
the non-redundant points of $\cQ = \cl(\cP)$;
hence, it is the minimal system such that $\fullgen(\SK_\cQ) = \cQ$.
\qed
\end{delayedproof}


\begin{delayedproof}[Proposition~\ref{prop:multiple-mater}]
Let $\SK_F = \langle L_F, R_F, C_F, \emptyset \rangle \sseq \SK_\cQ$
be the skeleton of the face $F \sseq \cQ$, so that $\fullgen(\SK_F) = F$.
By definition of `$\fullgen$',
the points $\vect{p}, \vect{p}' \in \relint(F)$
can be obtained by combining the generators in $\SK_F$:
\begin{align*}
  \vect{p}
    &=  L_F \vect{\lambda} + R_F \vect{\rho} + C_F \vect{\gamma}, \\
  \vect{p}'
    &=  L_F \vect{\lambda}' + R_F \vect{\rho}' + C_F\vect{\gamma}',
\end{align*}
where
$\vect{\lambda}, \vect{\lambda}' \in \Rset^{\ell}$,
$\vect{\rho}, \vect{\rho}' \in \nonnegRset^{r}$,
$\vect{\gamma}, \vect{\gamma}' \in \nonnegRset^{c}$,
$\sum_{i=1}^{c} \gamma_i = \sum_{i=1}^{c} \gamma_i' = 1$
and, for each $i \in \{ 1, \dots, c \}$,
both $\gamma_i >0$ and $\gamma_i' > 0$.
Therefore,
\begin{align*}
  \relint(F)
    = \gen\bigl(\langle L_F, R_F, C_F, \{\vect{p}\} \rangle\bigr)
    = \gen\bigl(\langle L_F, R_F, C_F, \{\vect{p}'\} \rangle\bigr).
\end{align*}
Hence we have shown that, in order to generate $\relint(F) \sseq \cP$,
point $\vect{p} \in P$ can be replaced by
any other point $\vect{p}' \in \relint(F)$.

By definition of `$\gen$',
the contribution of $\vect{p} \in P$ is to generate
the sets $\relint(F') \sseq \cP$, where $F' \in \NNCFaces$
is such that $\relint(F) \sseq F'$
(i.e., all the faces of $\cP$ containing $\relint(F)$).
It follows that $\vect{p} \in P$ can be substituted
by any other point $\vect{p}' \in \relint(F)$,
obtaining the same polyhedron.
\qed
\end{delayedproof}


\begin{delayedproof}[Proposition~\ref{prop:Galois-connection}]
Let $\SK$ be the skeleton of the polyhedron $\cP \in \Pset_n$,
$\cQ = \cl(\cP)$ and $\NSset$ be the corresponding set of supports.
In order to prove that $(\alpha_\SK, \gamma_\SK)$
is a Galois connection between $\wp(\cQ)$ and $\wpup(\NSset)$,
we will show that
`$\alpha_{\SK}$' and `$\gamma_{\SK}$' are monotonic,
`$\alpha_{\SK} \comp \gamma_{\SK}$' is reductive and
`$\gamma_{\SK} \comp \alpha_{\SK}$' is extensive;
the result will thus follow from~\cite[Theorem~5.3.0.4]{CousotC79}.

The monotonicity of both `$\alpha_{\SK}$' and `$\gamma_{\SK}$'
follows trivially from Definition~\ref{def:alpha-gamma}.

Consider $\NS \in \wpup(\NSset)$.
Note that, for each $\ns \in \NS$, there exists a face $F \in \CFaces$
such that $\ns = \SK_F$, so that $\fullgen(\ns) = F$.
Therefore,
\begin{align*}
  \alpha_{\SK}&\bigl(\gamma_{\SK}(\NS)\bigr) \\
    &\textrm{[by definition of $\gamma_\SK$]} \\
    &= \alpha_{\SK}
         \Bigl(
           \bigcup \bigl\{\,
             \relint(\fullgen(\ns))
           \bigm|
             \ns \in \NS
           \,\bigr\}
         \Bigr) \\
    &= \alpha_{\SK}
         \Bigl(
           \bigcup \bigl\{\,
             \relint(F)
           \bigm|
             F = \fullgen(\ns) \in \CFaces, \ns \in \NS
           \,\bigr\}
         \Bigr) \\
   &\textrm{[by definition of $\alpha_\SK$]} \\
   &= \bigcup
        \bigl\{\,
          \upcl \ns
        \bigm|
          \exists \vect{p} \in \relint(F),
          F = \fullgen(\ns) \in \CFaces, \ns \in \NS
        \,\bigr\} \\
   &= \bigcup \{\, \upcl \ns \mid \ns \in \NS \,\} \\
   &\textrm{[since $\NS$ is an upward closed set]} \\
   &= \NS.
\end{align*}
Hence, `$\alpha_{\SK} \comp \gamma_{\SK}$' is the identity function,
which implies that it is reductive.

In order to show that `$\gamma_{\SK} \comp \alpha_{\SK}$' is extensive,
let $S \sseq \cQ$.
Note that for each point $\vect{p} \in S$,
there exists a face $F \in \CFaces$ such that $\vect{p} \in \relint(F)$.
Hence:
\begin{align*}
  \gamma_{\SK}&\bigl(\alpha_{\SK}(S)\bigr) \\
    &\textrm{[by definition of $\alpha_\SK$]} \\
    &= \gamma_\SK\Bigl(
         \bigcup
           \bigl\{\,
             \upcl \SK_F
           \bigm|
             \exists \vect{p} \in S, F \in \CFaces
               \st \vect{p} \in \relint(F)
           \,\bigr\}
         \Bigr) \\
    &= \gamma_\SK\Bigl(
         \bigcup
           \bigl\{\,
             \ns
           \bigm|
             \exists \vect{p} \in S, F \in \CFaces
               \st \vect{p} \in \relint(F),
             \ns \in \upcl \SK_F
           \,\bigr\}
         \Bigr) \\
    &\textrm{[by definition of $\gamma_\SK$]} \\
    &= \bigcup
         \sset{
           \relint\bigl(\fullgen(\ns)\bigr)
         }{
           \exists \vect{p} \in S, F \in \CFaces
             \st \vect{p} \in \relint(F), \\
           \ns \in \upcl \SK_F
         } \\
    &\Sseq
      \bigcup
         \bigl\{\,
           \relint(F)
         \bigm|
           \exists \vect{p} \in S, F \in \CFaces
             \st \vect{p} \in \relint(F)
         \,\bigr\} \\
    &\Sseq
       \bigcup
         \bigl\{\,
           \vect{p} \in S
         \bigm|
           \exists F \in \CFaces \st \vect{p} \in \relint(F)
         \,\bigr\} \\
    &= S.
\end{align*}
\qed
\end{delayedproof}


\begin{delayedproof}[Proposition~\ref{prop:fixpoint}]
Applying `$\gamma_{\SK}\circ\alpha_{\SK}$' to the set of points $P$
we obtain:
\begin{equation}
\label{eq:gamma-alpha-P}
\gamma_{\SK}\bigl(\alpha_{\SK}(P)\bigr)
  = \bigcup
      \sset{
        \relint\bigl(\fullgen(\SK_F)\bigr)
      }{
        \exists \vect{p}' \in P, F' \in \CFaces \st \\
        \vect{p}' \in \relint(F'), \\
        \SK_F \in \upcl \SK_{F'}
      }.
\end{equation}

By definition of function `$\gen$',
a face is included in the polyhedron $\cP$
if and only if it contains a point in $P$.
In particular,
letting $\NNCFaces' = \NNCFaces \setdiff \{ \emptyset \}$,
this holds for the minimal faces in $\NNCFaces'$;
these are the atoms of the lattice $\cl(\NNCFaces)$,
which is a sublattice of $\CFaces$.
For these atoms $A \in \cl(\NNCFaces)$,
we have $A = \relint(A)$; hence
\begin{equation}
\label{eq:atoms-have-points}
  \forall A \textrm{ atom of } \cl(\NNCFaces) \itc
    \exists \vect{p} \in P \st \vect{p} \in \relint(A).
\end{equation}
Moreover, since $\cl(\NNCFaces')$ is an upward closed set,
we have:
\begin{equation}
\label{eq:atoms-are-enough}
  \forall F' \in \cl(\NNCFaces') \itc
    \exists A \textrm{ atom of } \cl(\NNCFaces) \st
      \SK_A \sseq \SK_{F'}.
\end{equation}
Therefore, we have the following chain of equations:
\begin{align}
  \cP
    &= \bigcup
         \bigl\{\,
           \relint(F)
         \bigm|
           F \in \NNCFaces'
         \,\bigr\}
       \nonumber \\
    &= \bigcup
         \bigl\{\,
           \relint\bigl(\fullgen(\SK_F)\bigr)
         \bigm|
           F \in \cl(\NNCFaces')
         \,\bigr\}
       \nonumber \\
    &\textrm{[by property~(\ref{eq:atoms-are-enough})]}
       \nonumber \\
    &= \bigcup
         \bigl\{\,
           \relint\bigl(\fullgen(\SK_F)\bigr)
         \bigm|
           \exists A \textrm{ atom of } \cl(\NNCFaces) \st \SK_A \sseq \SK_F
         \,\bigr\}
       \nonumber \\
    &\textrm{[by property~(\ref{eq:atoms-have-points})]}
       \nonumber \\
    &= \bigcup
         \sset{
           \relint\bigl(\fullgen(\SK_F)\bigr)
         }{
           \exists \vect{p} \in P, A \textrm{ atom of } \cl(\NNCFaces) \st \\
           \vect{p} \in \relint(A), \SK_F \in \upcl \SK_A
         }.
  \label{eq:big-set}
\end{align}
We now show that~(\ref{eq:big-set}) is equivalent to~(\ref{eq:gamma-alpha-P}).
The inclusion
(\ref{eq:big-set}) $\subseteq$ (\ref{eq:gamma-alpha-P})
follows by simply taking $F' = A$;
the other inclusion
(\ref{eq:big-set}) $\supseteq$ (\ref{eq:gamma-alpha-P})
follows by applying property~(\ref{eq:atoms-are-enough})
while also observing that, since $\vect{p}' \in \relint(F')$,
then $F' \in \cl(\NNCFaces)$.
\qed
\end{delayedproof}
